\documentclass[aps,prb,twocolumn,floatfix,superscriptaddress,longbibliography,showpacs]{revtex4-1}

\usepackage[utf8]{inputenc}
\usepackage{graphicx}
\usepackage{dcolumn}
\usepackage{bm}
\usepackage{amsfonts}
\usepackage{amsmath}
\usepackage{amssymb}
\usepackage{braket}
\usepackage{color,soul}
\usepackage{wasysym}
\usepackage{mathrsfs}
\usepackage{times}
\usepackage[dvipsnames,svgnames,table]{xcolor}
\usepackage{ulem}
\usepackage{hyperref}
\usepackage{fancyhdr}
\usepackage[english]{babel}
\usepackage{hyperref}
\usepackage{longtable}
\usepackage{multirow}

\hypersetup{
    unicode=true,          
    pdftoolbar=true,        
    pdfmenubar=true,        
    pdffitwindow=false,     
    pdfstartview={FitH},    
    pdfauthor={LEFFT},     	
    colorlinks=true,       	
    linkcolor=NavyBlue,  
    citecolor=Maroon,       
    filecolor=NavyBlue,      
    urlcolor=NavyBlue       
}

\newcommand{\ams}{\text{ }\textup{\AA}}
\newcommand{\tf}[1]{\mathrm{#1}}
\newcommand{\hbm}[1]{\hat{\bm{#1}}}

\begin{document}

\title{Floquet boundary states in AB-stacked graphite}

\author{Hern\'an L. Calvo}
\affiliation{Instituto de F\'isica Enrique Gaviola (CONICET) and FaMAF, Universidad Nacional de C\'ordoba, Argentina}
\affiliation{Departamento de F\'isica, Universidad Nacional de R\'io Cuarto, Ruta 36, Km 601, 5800 R\'io Cuarto, Argentina}

\author{Jose E. Barrios Vargas}
\affiliation{Departamento de F\'isica y Qu\'imica Te\'orica, Facultad de Qu\'imica, UNAM, 04510 M\'exico City, M\'exico}

\author{Luis E. F. Foa Torres}
\affiliation{Departamento de F\'isica, Facultad de Ciencias F\'isicas y Matem\'aticas, Universidad de Chile, Santiago, Chile}

\begin{abstract}
We report on the effect of laser illumination with circularly polarized light on the electronic structure of AB-stacked graphite samples. By using Floquet theory in combination with Green's function techniques, we find that the polarized light induces bandgap openings at the Floquet zone edge $\hbar\Omega/2$, bridged by chiral boundary states. These states propagate mainly along the borders of the constituting layers as evidenced by the time-averaged local density of states and the probability current density in several geometries. Semianalytic calculations of the Chern number suggest that these states are of topological nature, similar to those found in illuminated 2D samples like monolayer and bilayer graphene. These states are promising candidates for the realization of a three-dimensional version of the quantum Hall effect for Floquet systems.
\end{abstract}

\maketitle

\section{Introduction}\label{sec:intro}

Condensed matter physics has provided one of the most fertile and captivating grounds for discoveries over the last few decades:~\cite{martin2019} from two-dimensional materials,~\cite{novoselov2004} which were thought to be impossible to exist in nature, to new topological phases of matter~\cite{noauthor2016,castelvecchi2017,hasan2010,xiao2010,asboth2016} which have completely reshaped our understanding of old concepts. The use of light has been an instrumental cornerstone in this adventure, being one of the prime tools for unveiling a material or device properties.~\cite{xu2015,hasan2017} However, beyond this already important role, a new research front aims at using light in an active fashion to actually change the response of a material,~\cite{oka2009,lindner2011} by opening a gap~\cite{syzranov2008,lopez-rodriguez2008,oka2009,calvo2011,wang2013} or even endowing a material with topological states~\cite{oka2009,lindner2011,kitagawa2011,gomez-leon2013,piskunow2014,rudner2019a} or a spontaneous orbital magnetization.~\cite{rudner2019b}

Experiments have successfully confirmed the possibility of creating and tuning hybrid electron-photon states,~\cite{wang2013,mahmood2016} also called Floquet-Bloch states, and also the generation of a laser-induced Hall effect in graphene.~\cite{mciver2018} The name Floquet here is used because the prevalent theory for this type of driven systems: the Floquet theory,~\cite{shirley1965,sambe1973,kohler2005,platero2004} from which the spectrum, effective Hamiltonians,~\cite{sandovalsantana2019,vogl2019} a map of the topological invariants~\cite{rudner2013,piskunow2015} and transport properties~\cite{kundu2013,foa_torres2014,rodriguez-mena2019} can be computed. It is also worthy to mention that this Floquet picture can be ported, with small changes, to phonon-induced states as in Refs.~[\onlinecite{calvo2018}] and [\onlinecite{chaudhary2019}]. Most of the attention has been devoted to illuminating two-dimensional materials, including graphene,~\cite{oka2009,dehghani2014,dehghani2015,usaj2014} germanene,~\cite{tahir2016} silicene,~\cite{ezawa2013} MoS$_2$,~\cite{huaman2019} and manufactured systems like periodic arrangements of quantum rings.~\cite{kozin2018} More recently, the interest in Floquet engineering three-dimensional materials such as three-dimensional topological insulators,~\cite{fregoso2014,calvo2015} Weyl semimetals~\cite{gonzalez2016,zhang2016} or Dirac materials~\cite{islam2019} has been growing.

Here we focus on laser-illuminated graphite. In contrast with most three-dimensional crystals, graphite has a hierarchical structure of weakly coupled layers making it an archetypal system for learning on the way from two to three dimensions. In two-dimensions, circularly polarized radiation leads to bandgap openings and Floquet edge states that bridge the gap.~\cite{oka2009,dehghani2014,dehghani2015,piskunow2014,usaj2014} These topological Floquet edge states are akin to those found in Chern insulators or in the integer quantum Hall effect, they are robust and chiral.~\cite{piskunow2014} By analogy with the physics of the quantum Hall effect which was discovered in two dimensions~\cite{von_klitzing1980,thouless1982} and which has been predicted to be possible in three dimensions,~\cite{halperin1987} a prediction which has not been verified until very recently,~\cite{tang2019} one might then wonder what happens in three-dimensions with the laser-induced states. In this paper, we show that for graphite there are also laser-induced bandgaps at $\pm \hbar\Omega/2$ which turn out to be bridged by surface states. Our calculations, which are based on Green's functions techniques combined with Floquet theory, show that these surface states are chiral, have a topological nature and can form a band of chiral states bridging the bulk gap. In the following we introduce our model, followed by an analysis of bulk graphite, and finite samples with emphasis on the surface states and the associated currents. 

\section{Hamiltonian model and Floquet space solution scheme}\label{sec:model}

Let us introduce our model for graphite under circularly polarized laser illumination. We consider graphene layers in graphite with AB stacking, and we follow Ref.~[\onlinecite{dresselhaus1981}] for the tight-binding parameters obtained in the static case (see below). We consider a 
tight-binding description for graphite given by the generic Hamiltonian
\begin{equation}
\hat{\mathcal{H}} = \sum_{\bm{r},\bm{r}'} \gamma_{\bm{r},\bm{r}'} \ket{\bm{r}}\!\bra{\bm{r}'},
\label{eq:H}
\end{equation}
where $\bm{r}$ and $\bm{r}'$ denote the positions of the carbon atoms in the lattice, such that the states $\ket{\bm{r}}$ form a real-space basis. 
Under this notation, the sum runs over sites connected by the hopping amplitudes $\gamma_{\bm{r},\bm{r}'}$, and it also includes the on-site energies through $\gamma_{\bm{r},\bm{r}} = \epsilon_{\bm{r}}$.

The laser field is included within a semiclassical approximation as a time-dependent term in the Hamiltonian. The time-periodic electric field $\bm{E}(\bm{r},t)$, with
period $T = 2\pi/\Omega$, is included through the gauge $\bm{E} = - \partial_t \bm{A}$, where the 
vector potential takes the form $\bm{A}(\bm{r},t) = \tf{Re} [\bm{A}_0 e^{i\Omega(z/c - t)}]$, such that its direction of propagation points 
perpendicular to the graphene layers, defined in the $xy$ planes. As a consequence, in three dimensional samples there is a variation of the wave along the $z$ direction due to the phase factor $\Omega z/c$ in $\bm{A}(\bm{r},t)$. This would become appreciable in samples with transversal lengths higher than $L_z \sim 0.1 \lambda$, with $\lambda = 2\pi c/\Omega$ the laser's wavelength. For laser frequencies near the infrared region ($\hbar\Omega \sim 2$ eV) this implies $L_z \sim 620\ams$, which in graphite means a number of $\sim \! 185$ transversal layers. As we will assume smaller values for $L_z$, the $z$ dependence in the vector potential can be neglected in a first approximation. We will work with circularly polarized light, by  taking $\bm{A}_0 = A_0 (1,i\tau,0)$, with $\tau = \pm 1$ the handedness of the polarized light.~\footnote{Throughout the manuscript we will take $\tau = 1$, unless otherwise indicated.} By means of Peierls' substitution, the vector potential enters in Eq.~(\ref{eq:H}) by adding a time-dependent phase in the hopping amplitudes, namely,
\begin{equation}
\gamma_{\bm{r},\bm{r}'} \xrightarrow{\tf{laser}} g_{\bm{r},\bm{r}'}(t) = \gamma_{\bm{r},\bm{r}'} \exp\left[ i \frac{2\pi}{\Phi_0} \int_{\bm{r}'}^{\bm{r}} \tf{d}\bm{\ell} \cdot \bm{A}(t) \right],
\label{eq:hopp1}
\end{equation}
with $\Phi_0$ the magnetic flux quantum and the line integral taken over the straight path connecting sites $\bm{r}'$ and $\bm{r}$. Given the specific form of 
the vector potential, the time-dependent hopping terms entering in the Hamiltonian are given by:
\begin{equation}
g_{\bm{r},\bm{r}'}(t) = \gamma_{\bm{r},\bm{r}'} \sum_{n=-\infty}^\infty i^n \mathcal{J}_n(\zeta_{\bm{r},\bm{r}'}) e^{in(\Omega t-\phi_{\bm{r},\bm{r}'})},
\label{eq:hopp2}
\end{equation}
where we used the Jacobi-Anger expansion for future convenience. In this expression, $\mathcal{J}_n(\zeta_{\bm{r},\bm{r}'})$ represents the $n$-th Bessel function of the first kind, and the adimensional variable $\zeta_{\bm{r},\bm{r}'} = 2\pi A_0 |\bm{r}-\bm{r}'|\sin \theta_{\bm{r},\bm{r}'}/\Phi_0$ quantifies the strength of the laser along the carbon bond, characterized by $\bm{r}-\bm{r}' = |\bm{r}-\bm{r}'| (\sin\theta_{\bm{r},\bm{r}'}\cos\phi_{\bm{r},\bm{r}'},\,\sin \theta_{\bm{r},\bm{r}'} \sin \phi_{\bm{r},\bm{r}'},\,\cos \theta_{\bm{r},\bm{r}'})$.  

\subsection{Floquet theory} \label{sec:floquet}

In this section we introduce the basics of Floquet theory as used later in this paper. The readers already acquainted with the technical details or focused on the results rather than the techniques may skip this in a first reading.

According to Floquet theory,~\cite{shirley1965,sambe1973,kohler2005,platero2004} there is a full set of solutions to the time-dependent Schr\"odinger equation (TDSE) of the form 
$\ket{\psi(t)} = e^{-i\epsilon t/\hbar}\ket{\phi(t)}$, where the Floquet state $\ket{\phi(t)}$ presents the same 
periodicity of the Hamiltonian, i.e. $\ket{\phi(t+T)} = \ket{\phi(t)}$. By replacing this ansatz in the TDSE one obtains
\begin{equation}
\hat{\mathcal{H}}_\tf{F} \ket{\phi(t)} = \epsilon \ket{\phi(t)},
\label{eq:Floquet-eq}
\end{equation} 
where $\hat{\mathcal{H}}_\tf{F}=\hat{\mathcal{H}}(t)-i\hbar\partial_t$ is the so-called Floquet Hamiltonian and $\epsilon$ its associated
quasienergy. The great advantage of Floquet theory is that $\hat{\mathcal{H}}_\tf{F}$ can be reduced to a time-independent matrix when 
described in the product space (also called Floquet space) $\mathcal{F} = \mathcal{R}\otimes\mathcal{T}$, with $\mathcal{R}$ the usual Hilbert space and $\mathcal{T}$ 
the space of time-periodic functions, spanned by the set of orthonormal vectors $\braket{t|n} = e^{in\Omega t}$, with $n$ an integer number. Working 
within the local space representation, a suitable basis for $\mathcal{F}$ is given by the product states 
$\ket{\bm{r},n} = \ket{\bm{r}}\otimes\ket{n}$, representing the lattice site $\bm{r}$ and the Fourier replica $n$, together with the inner 
product rule
\begin{equation}
\braket{\bm{r},n|\bm{r}',m} = \int_0^T \frac{\tf{d}t}{T} e^{i(m-n)\Omega t} \braket{\bm{r}|\bm{r}'} = \delta_{\bm{r},\bm{r}'}\delta_{n,m}.
\end{equation}
On this basis, the Floquet states in Eq.~(\ref{eq:Floquet-eq}) can be computed as
\begin{equation}
\ket{\phi(t)} = \sum_n e^{i n \Omega t} \ket{\phi_n} \xrightarrow{\,\mathcal{F}\,} \ket{\phi} = \sum_{\bm{r},n} \phi_n(\bm{r}) \ket{\bm{r},n},
\end{equation}
with $\phi_n(\bm{r}) = \braket{\bm{r},n|\phi}$ the amplitude of the Floquet state at site $\bm{r}$ and replica $n$. Importantly, the
matrix elements of the Floquet Hamiltonian $[H_\tf{F}]_{\bm{r},\bm{r}'}^{n,m} = \bra{\bm{r},n} \hat{\mathcal{H}}_\tf{F} 
\ket{\bm{r}',m}$ are in this basis
\begin{equation}
[H_\tf{F}]_{\bm{r},\bm{r}'}^{n,m} = \int_0^T \frac{\tf{d}t}{T} e^{i (m-n) \Omega t} H_{\bm{r},\bm{r}'}(t),
+n\hbar\Omega\delta_{\bm{r},\bm{r}'}\delta_{nm},
\label{eq:HF}
\end{equation}
where the inner product includes the average over one driving cycle, thus Eq.~(\ref{eq:Floquet-eq}) written in this composite space becomes a 
time-independent eigenvalue problem. Once the Floquet eigenstates $\ket{\phi}$ are obtained in $\mathcal{F}$, it is possible to return to the usual 
Hilbert space $\mathcal{R}$ and calculate the expectation value of any observable from the general solution $\ket{\psi(t)}$ of the TDSE. In particular, we are interested in the probability density $\hat{\rho}(\bm{r}) = \ket{\bm{r}}\!\bra{\bm{r}}$, whose time-averaged expectation value with respect to some 
eigenstate of the TDSE writes
\begin{equation}
\rho(\bm{r}) = \int_0^T \frac{\tf{d}t}{T} \braket{\hat{\rho}(\bm{r})} = \sum_{n} |\phi_n(\bm{r})|^2.
\label{eq:rho}
\end{equation}
We are also interested in the probability current density, defined as
$\hat{J}(\bm{r},t) = -i[\hat{\mathcal{H}}(t),\hat{\rho}(\bm{r})]/\hbar$. Its time-averaged expectation value can be written in terms of the Floquet Hamiltonian
\begin{equation}
J(\bm{r}) = \frac{2}{\hbar} \sum_{\bm{r}'} \sum_{n,m} \tf{Im} \{ \phi_n^\ast(\bm{r})[H_\tf{F}]_{\bm{r},\bm{r}'}^{n,m} \phi_m(\bm{r}') \}.
\end{equation}
Since the averaged probability current at site $\bm{r}$ needs to be zero due to probability conservation, we will use this quantity to check that 
there is no charge accumulation/loss after completing one period of the driving field. More interestingly, from this expression we can extract the 
bond current as~\cite{todorov2002} 
\begin{equation}
J(\bm{r},\bm{r}') = \frac{2}{\hbar} \sum_{n,m} \tf{Im} \{ \phi_n^\ast(\bm{r})[H_\tf{F}]_{\bm{r},\bm{r}'}^{n,m} \phi_m(\bm{r}') \},
\label{eq:current}
\end{equation}
which, as we will show later on, gives a clear picture on the chiral nature of the resulting eigenstates of the illuminated system. 

As we already mentioned, the periodic time-dependence enters in Eq.~(\ref{eq:hopp1}) as an additional phase that the electron picks up when it ``hops'' from site $\bm{r}'$ to site $\bm{r}$. The Floquet Hamiltonian can then be calculated from Eq.~(\ref{eq:HF}) as
\begin{equation}
[H_\tf{F}]_{\bm{r},\bm{r}'}^{n,m} = \gamma_{\bm{r},\bm{r}'}^{(m-n)} + n\hbar\Omega\delta_{\bm{r},\bm{r}'}\delta_{nm},
\end{equation}
where the hopping amplitudes are defined as Fourier components of the time-dependent ones appearing in Eq.~(\ref{eq:hopp2}), i.e.,
\begin{equation}
\gamma_{\bm{r},\bm{r}'}^{(n)} = \int_0^T \frac{\tf{d}t}{T} g_{\bm{r},\bm{r}'}(t)e^{i n\Omega t} = \gamma_{\bm{r},\bm{r}'} i^n \mathcal{J}_n(\zeta_{\bm{r},\bm{r}'}) e^{in\phi_{\bm{r},\bm{r}'}},
\end{equation}
and this can be \textit{interpreted} as the probability amplitude for the electron to hop from site $\bm{r}'$ to site $\bm{r}$, together with the 
absorption ($n>0$) or emission ($n<0$) of $|n|$ photons.

\begin{figure}[t]
\centering
\includegraphics[width=0.75\columnwidth]{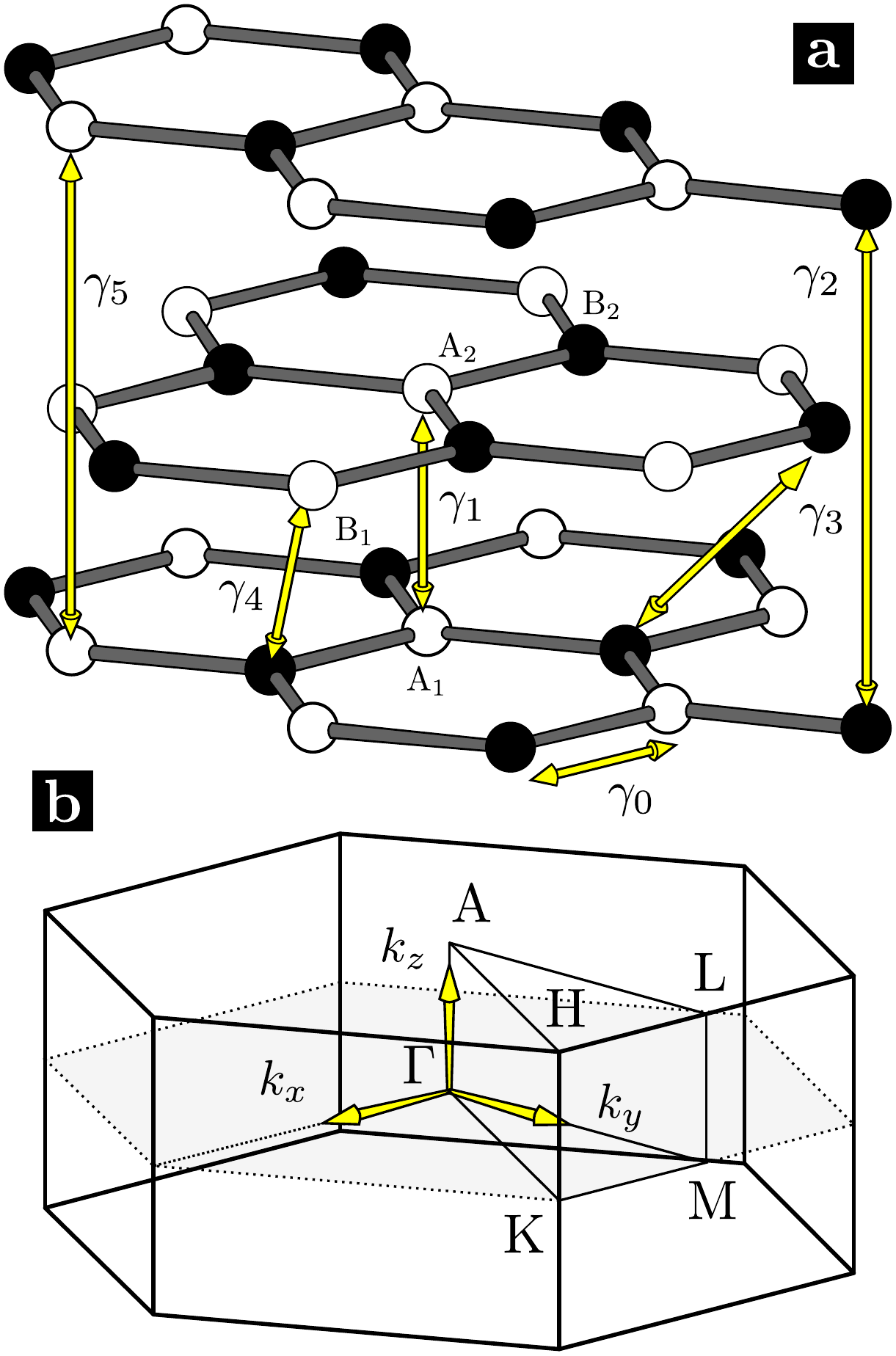}
\caption{Graphite with AB stacking. (a) Schematic view of three adjacent graphene layers. The hopping amplitudes are marked by yellow arrows 
(see text). (b) Graphite's first Brillouin zone with the high-symmetry points.}
\label{fig:1}
\end{figure}

So far we have not specified the tight-binding Hamiltonian of Eq.~(\ref{eq:H}), so the above discussion is somewhat general as far as the light propagates along the $z$ direction. In graphite with AB stacking, the unit cell contains four basis sites: A$_1$ and B$_1$ in the lower graphene layer (LL), A$_2$ and B$_2$ in the upper graphene layer (UL), see Fig.~\ref{fig:1}. The sites in the UL are displayed in such a way that the A$_2$ site is aligned with the A$_1$ site of the LL. This implies the following choice for the basis vectors in the unit cell:
\begin{align}
\tf{LL} &\rightarrow & \boldsymbol{\delta}_{\tf{A}_1} &= \left(0,0,0\right), & \boldsymbol{\delta}_{\tf{B}_1} &= \left(0,a_0,0\right), \\
\tf{UL} &\rightarrow & \boldsymbol{\delta}_{\tf{A}_2} &= \left(0,0,c_0\right), & \boldsymbol{\delta}_{\tf{B}_2} &= \left(\frac{\sqrt{3}a_0}{2},\frac{a_0}{2},c_0\right),
\end{align}
where $a_0 = 1.42\ams$ is the carbon-carbon distance in graphene and $c_0 = 3.35\ams$ is the separation between two adjacent layers. The graphite's 
Bravais lattice can then be described by the primitive vectors
\begin{align}
\bm{a}_1 &= \left( \frac{\sqrt{3}a_0}{2},\frac{3a_0}{2},0\right), \; \bm{a}_2 = \left(-\frac{\sqrt{3}a_0}{2},\frac{3a_0}{2},0\right) \nonumber \\
\bm{a}_3 &= \left(0,0,2c_0\right).
\label{eq:primitive}
\end{align}
The translational invariance along the three directions given by the primitive vectors allows us to decompose the static Hamiltonian in Eq.~(\ref{eq:H}) as the following operator representing the Bloch Hamiltonian:
\begin{equation}
\hat{\mathcal{H}}_{\bm{k}} = \sum_{\bm{R}} \hat{\mathcal{V}}_{\bm{R}} e^{i\bm{k}\cdot\bm{R}},
\label{eq:HB}
\end{equation}
where $\bm{R}$ denotes the position of the nearest-neighbor unit cells to the one placed at the origin. In the general description of the unit cell 
position through $\bm{R} = n_1\bm{a}_1+n_2\bm{a}_2+n_3\bm{a}_3$, with $n_i$ integer numbers, the lattice connectivity given by the hopping parameters 
determined in Ref.~[\onlinecite{dresselhaus1981}] implies that the possible values for $n_i$ in $\bm{R}$ are $n_i = \{-1,0,1\}$. To reconcile the notation, we notice that the hopping operator $\hat{\mathcal{V}}_{\bm{R}}$ represents the bonds going from site $\bm{r}' = \boldsymbol{\delta}(\bm{r}')$ to site $\bm{r} = \bm{R}+\boldsymbol{\delta}(\bm{r})$, where $\boldsymbol{\delta}(\bm{r})$ indicates the basis vector associated with $\bm{r}$.

In the case of graphite with AB stacking, we consider for the static case the following parameters:~\cite{dresselhaus1981} $\gamma_0 = 3.16$ eV connecting nearest-neighbor in-plane sites (A$_1$B$_1$ and A$_2$B$_2$), $\gamma_1 = 0.39$ eV for A$_1$A$_2$, $\gamma_2 = -0.02$ eV connecting B-sites (B$_1$B$_1$ and B$_2$B$_2$) between consecutive cells along $\bm{a}_3$, $\gamma_3 = 0.315$ eV for B$_1$B$_2$, $\gamma_4 = 0.044$ eV for A$_1$B$_2$ and B$_1$A$_2$, and $\gamma_5 = 0.038$ eV connecting A-sites (A$_1$A$_1$ and A$_2$A$_2$) between consecutive cells along $\bm{a}_3$. This can be easily understood, for example, by inspecting the matrix elements of the Bloch Hamiltonian with respect to the site basis $\{\ket{i}\}$, with $i=1,\dots,4$ for $(\tf{A}_1,\tf{B}_1,\tf{A}_2,\tf{B}_2)$:
\begin{widetext}
\begin{equation}
\bm{H}_{\bm{k}} = \left(\begin{matrix}
\epsilon_0+\Delta+\gamma_5 f_5	&	\gamma_0 f_1				&	\gamma_1 f_4					&	\gamma_4 f_2 f_4		\\
\gamma_0 f_1^*					&	\epsilon_0+\gamma_2 f_5		&	\gamma_4 f_1^* f_4				&	\gamma_3 f_3 f_4		\\
\gamma_1 f_4^*					&	\gamma_4 f_1 f_4^*			&	\epsilon_0+\Delta+\gamma_5 f_5	&	\gamma_0 f_2			\\
\gamma_4 f_2^* f_4^*			&	\gamma_3 f_3^* f_4^*		&	\gamma_0 f_2^*					&	\epsilon_0+\gamma_2 f_5
\end{matrix}\right),
\label{eq:bloch}
\end{equation}
\end{widetext}
where $\epsilon_0 = -0.024$ eV is the Fermi energy and $\Delta = -0.008$ eV is the energy shift between inequivalent carbon atoms. The functions $f_i = f_i(\bm{k})$ carry information about the directions in which the basis sites in the unit cell are connected with its neighbors, 
and are defined as:
\begin{align*}
f_1 &= 1+e^{i\bm{k}\cdot\bm{a}_1}+e^{i\bm{k}\cdot\bm{a}_2},				\\
f_2 &= 1+e^{i\bm{k}\cdot\bm{a}_1}+e^{i\bm{k}\cdot(\bm{a}_1-\bm{a}_2)} = e^{i\bm{k}\cdot\bm{a}_1} f_1^*,	\\
f_3 &= 1+e^{-i\bm{k}\cdot\bm{a}_2}+e^{i\bm{k}\cdot(\bm{a}_1-\bm{a}_2)} = e^{-i\bm{k}\cdot\bm{a}_2} f_1, \\
f_4 &= 1+e^{i\bm{k}\cdot\bm{a}_3},\\
f_5 &= 2 \cos (\bm{k}\cdot \bm{a}_3).
\end{align*}
So, for example, in the matrix element $[\bm{H}_{\bm{k}}]_{12}$, the function $f_1$ accounts for the intracell connection 
$\boldsymbol{\delta}_{\tf{B}_1} \rightarrow \boldsymbol{\delta}_{\tf{A}_1}$ and the intercell connections $\boldsymbol{\delta}_{\tf{B}_1} \rightarrow \bm{a}_1+\boldsymbol{\delta}_{\tf{A}_1}$ and $\boldsymbol{\delta}_{\tf{B}_1} \rightarrow \bm{a}_2+\boldsymbol{\delta}_{\tf{A}_1}$. Similarly, $f_2$ in $[\bm{H}_{\bm{k}}]_{34}$ takes into account those bonds connecting $\boldsymbol{\delta}_{\tf{B}_2}$ with $\boldsymbol{\delta}_{\tf{A}_2}$. With this notation, it is clear that the coupling between different graphene layers enters in
the $2 \times 2$ off-diagonal blocks, which are modulated by either $f_4$ or $f_4^*$. In the diagonal blocks, on the other hand, there are
in-plane connections given by $\gamma_0$ and on-site energy corrections due to the coupling to neighbor cells along $\bm{a}_3$. Additionally, the difference in the involved directions given by $f_1$ in the LL and $f_2$ in the UL, 
respectively, comes from the choice of the unit cell basis sites. Notice, in particular, that $|f_1|=|f_2|=|f_3|$.

\begin{figure*}[th]
\centering
\includegraphics[width=\textwidth]{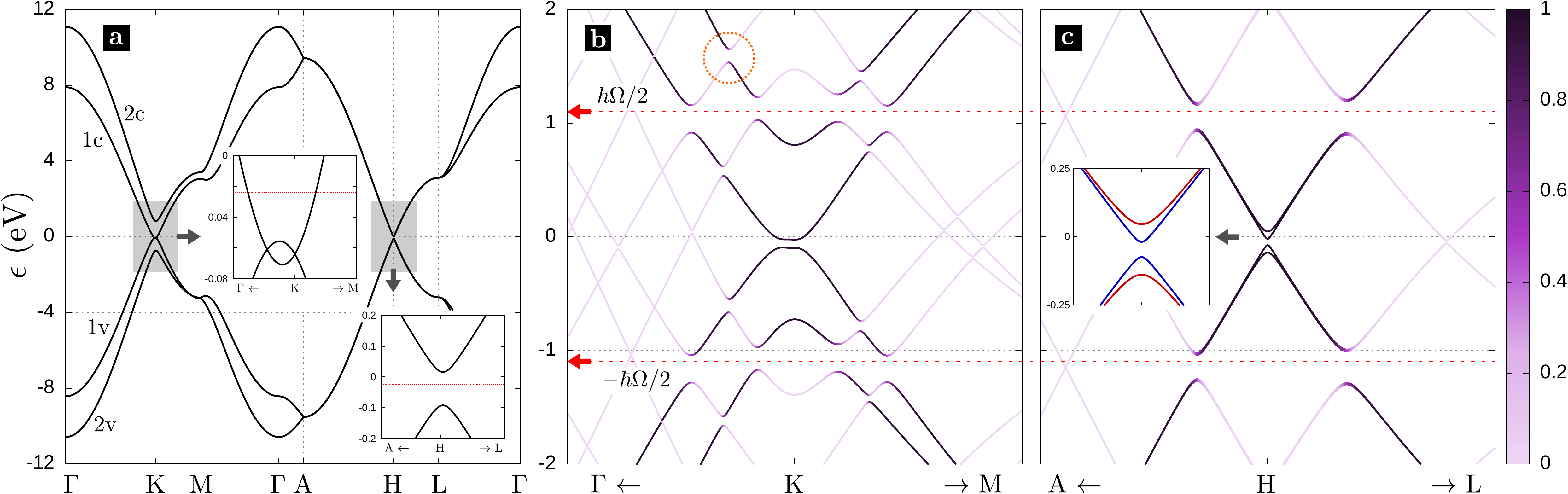}
\caption{Bulk graphite's dispersion relations. (a) Static case where no laser is applied. The labels in the $\bm{k}$-axis (horizontal) correspond to the symmetry points depicted in Fig.~\ref{fig:1}(b). The valence bands are labeled as 1v and 2v, while the conduction bands are labeled as 1c and 2c. The insets are zoom regions around the K and H symmetry points and the Fermi energy is depicted in dotted red. (b) and (c) are zoom regions around the K and H symmetry points [gray shaded rectangles in (a)], respectively, for circularly polarized light with $\zeta_0 = 0.04 \ams^{-1} \times 1.42 \ams = 0.0568$ and $\hbar\Omega = 2.2$ eV. The inset in (c) is a zoom around $\epsilon = 0$ and shows the LL-bands in solid red and the UL-bands in solid blue (see text). The colorscale represents the weight of the $\bm{k}$-states on the zeroth Fourier replica, according to Eq.~(\ref{eq:weight}).}
\label{fig:2}
\end{figure*} 

If we now turn on the laser, one should notice that the vector potential does not break translational invariance, so it is possible to combine Eq.~(\ref{eq:HB}) with Eq.~(\ref{eq:HF}) by introducing a superindex $(m-n)$ in the Bloch Hamiltonian which 
accounts for the replicas $m \rightarrow n$ it connects. This implies that all hoppings belonging to $\hat{\mathcal{H}}_{\bm{k}}^{(m-n)}$ 
need to be transformed as $\gamma_{\bm{r},\bm{r}'} \rightarrow \gamma_{\bm{r},\bm{r}'}^{(m-n)}$, and we obtain the following structure
\begin{equation}
\hat{\mathcal{H}}_{\tf{F},\bm{k}} = \sum_{n,m} \left[ \hat{\mathcal{H}}_{\bm{k}}^{(m-n)} + n\hbar\Omega \hat{\mathcal{I}} \delta_{n,m} \right] \otimes \ket{n}\!\bra{m},
\label{eq:HamF}
\end{equation}
for the Floquet-Bloch Hamiltonian, defined in the $\mathcal{F}$-space. Here $\hat{\mathcal{I}}$ represents the identity operator in the reduced space of the unit cell and $\ket{n}$ corresponds to the Fourier replica $n$.

Summarizing, the construction of $\hat{\mathcal{H}}_{\tf{F},\bm{k}}$ follows 
two simple steps: (1) the identification of the static Bloch Hamiltonian of Eq.~(\ref{eq:HB}), and (2) the Fourier decomposition of all their matrix 
elements once the laser has been incorporated. Notice that there is, however, a subtlety in going from step 1 to step 2: as the time-dependent hopping 
phases [cf. Eq.~(\ref{eq:hopp2})] depend on both the magnitude and direction of the bond connecting sites $\bm{r}'$ and $\bm{r}$, this information 
needs to be given in step 1 even if in the static case such a dependence is not present.

Following the above steps, the matrix elements of the Floquet-Bloch Hamiltonian for bulk graphite can be compactly written in terms of the hopping 
amplitudes between the different basis sites $i,j = \{\tf{A}_1,\tf{B}_1,\tf{A}_2,\tf{B}_2\}$ as
\begin{equation}
[H_{\tf{F},\bm{k}}]_{i,j}^{n,m} =  \sum_{\bm{R}} \gamma_{\bm{R}+\boldsymbol{\delta}_i,\boldsymbol{\delta}_j}^{(n-m)} e^{i\bm{k}\cdot\bm{R}} + 
n\hbar\Omega\delta_{i,j}\delta_{n,m}. \label{eq:bloch-floquet}
\end{equation}
Notice that not all lattice vectors $\bm{R}$ contribute to the sum on the right hand side, as we assume some finite range for the allowed hopping parameters.

\section{Illuminated bulk graphite}\label{sec:bulk}

The purpose of this section is to give an explicit calculation of the Floquet Hamiltonian in illuminated graphite, such that the role of the laser 
field is evidenced as modifications in the band structure of the static material. This will allow us to identify, in turn, the band crossing 
regions where boundary states induced by the laser may appear.  

As starting point, in Fig.~\ref{fig:2}(a), we show the dispersion relation for bulk graphite in the absence of laser illumination. We can see how the highest valence (1v) and the lowest conduction (1c) bands cross at the K symmetry point. These bands are quadratic in shape (a reminiscence of bilayer graphene's band structure), and cross each other at two different points: one of them along the $\Gamma$-K path while the other exactly at the K-point (see inset).~\cite{cheng2015} The breaking of the electron-hole (e-h) symmetry is clearly visible along the whole spectrum and is produced by the hoppings $\gamma_2$, $\gamma_4$, and $\gamma_5$. Along the A-H-L path [top face of the Brillouin zone in Fig.~\ref{fig:1}(b)] the energy bands become doubly degenerate. Inspecting the Bloch Hamiltonian in Eq.~(\ref{eq:bloch}), this band degeneracy can be easily understood since $f_4$ becomes exactly zero, meaning that the layers are completely decoupled along this path and, in addition, $|f_1|=|f_2|$. Exactly at the H point, there is a gap $\Delta\epsilon \simeq 124$ meV due to $\gamma_2$, $\gamma_5$ and $\Delta$.

For illuminated graphite, one should notice that an infinite number of replicas develop in the quasienergy spectrum associated with the Floquet-Bloch 
Hamiltonian. We are, however, interested in the changes that the laser field produces on the static spectrum shown in Fig.~\ref{fig:2}(a). A 
convenient way to visualize this is to use a colorscale that represents the weight of the $\bm{k}$-eigenstates on the $n=0$ Fourier replica, i.e.
\begin{equation}
\bar{w}_{\bm{k}} = \sum_{\bm{r}} |\phi_{\bm{k},0}(\bm{r})|^2,
\label{eq:weight}
\end{equation}
where the sum runs over the basis sites composing the unit cell, i.e. $\bm{r} = \{\boldsymbol{\delta}_{\tf{A}_1},\boldsymbol{\delta}_{\tf{B}_1},\boldsymbol{\delta}_{\tf{A}_2},\boldsymbol{\delta}_{\tf{B}_2}\}$. Comparing the above expression 
with Eq.~(\ref{eq:rho}), $\bar{w}_{\bm{k}}$ represents the fraction of the (time-averaged) probability density which is distributed along the $n=0$ 
replica. Notice that in the static case $\bar{w}_{\bm{k}} = \sum_{\bm{r}} \rho_{\bm{k}}(\bm{r}) = 1$ since no other replicas are involved. We set the strenght of the laser through $\zeta_0 = 2\pi A_0 a_0/\Phi_0 = 0.0568$, such that $\zeta_{\bm{r},\bm{r}'} = \zeta_0 |\bm{r}-\bm{r}'| \sin\theta_{\bm{r},\bm{r}'}/a_0$, and the frequency as $\hbar\Omega = 2.2$ eV. In this regime, no strong modifications of the entire 
band structure are expected and one can, in turn, safely truncate the full Floquet space by taking an adequate number of replicas such that the 
observed spectrum converges. For the chosen parameters, appreciable changes induced by the laser only appear around the K and H symmetry points 
where the bands come close to each other, so we can focus in the gray shaded rectangles of Fig.~\ref{fig:2}(a). This is plotted in Fig.~\ref{fig:2}(b) in the vicinity of the K point along the path $\Gamma$-K-M and in Fig.~\ref{fig:2}(c) for the vicinity of the H point along the path A-H-L, respectively. The main features in these plots are the bandgap openings that appear around the boundaries of the Floquet zone (FZB), defined at $\epsilon = \pm \hbar \Omega/2$. Although not clearly visible, there is also a bandgap opening around the center of the Floquet zone (FZC) at $\epsilon = 0$. The large difference in the magnitude of the two gaps obeys a simple reason: the gap in the FZB region depends linearly on the laser's strength, while for the FZC gap such a dependence is quadratic.~\cite{calvo2011}

As it happens in two-dimensional samples with circularly polarized light,~\cite{oka2009,calvo2011} the bandgap openings are a known consequence of the
breaking of the time-reversal symmetry, which in this case extends to three-dimensional graphite. In Fig.~\ref{fig:2}(b), we can also distiguish some 
avoided crossings above and below the FZB gaps, between different e-h band partners. Take for example the one marked by the dotted circle, which corresponds to the crossing between the 2c--0 and the 1v--1 bands, where ``--$n$'' means that it belongs to the $n$th-Fourier replica in the limit 
$\zeta_0 \rightarrow 0$. We can see, however, that this is not a fully developed gap since for that energy range the 1c--0 band (to the left) is barely affected by the laser. 

Interestingly, in Fig.~\ref{fig:2}(c), the band degeneracy observed along the A-H-L trajectory in Fig.~\ref{fig:2}(a) for the static case is removed by the laser (see inset). Although the lower and the upper layers are still decoupled, the combination of the broken sublattice symmetry, due to the on-site energies, i.e.,
\begin{align*}
\epsilon_{\tf{A}_1} &= \epsilon_{\tf{A}_2} = \epsilon_0 + \Delta - 2\gamma_5,\\
\epsilon_{\tf{B}_1} &= \epsilon_{\tf{B}_2} = \epsilon_0 - 2\gamma_2,
\end{align*}
together with the handedness of the circularly polarized waves, allows one to distinguish between LLs and ULs, since these are mirror images of each other. This is depicted in the inset of Fig.~\ref{fig:2}(c), where we use red for the LL-bands and blue for the UL-bands. If we change the handedness of the laser field, then the bands behavior is indeed inverted (i.e., ``red becomes blue'' and viceversa), as expected from the $z \rightarrow -z$ inversion operation.

\begin{figure*}[th]
\centering
\includegraphics[width=\textwidth]{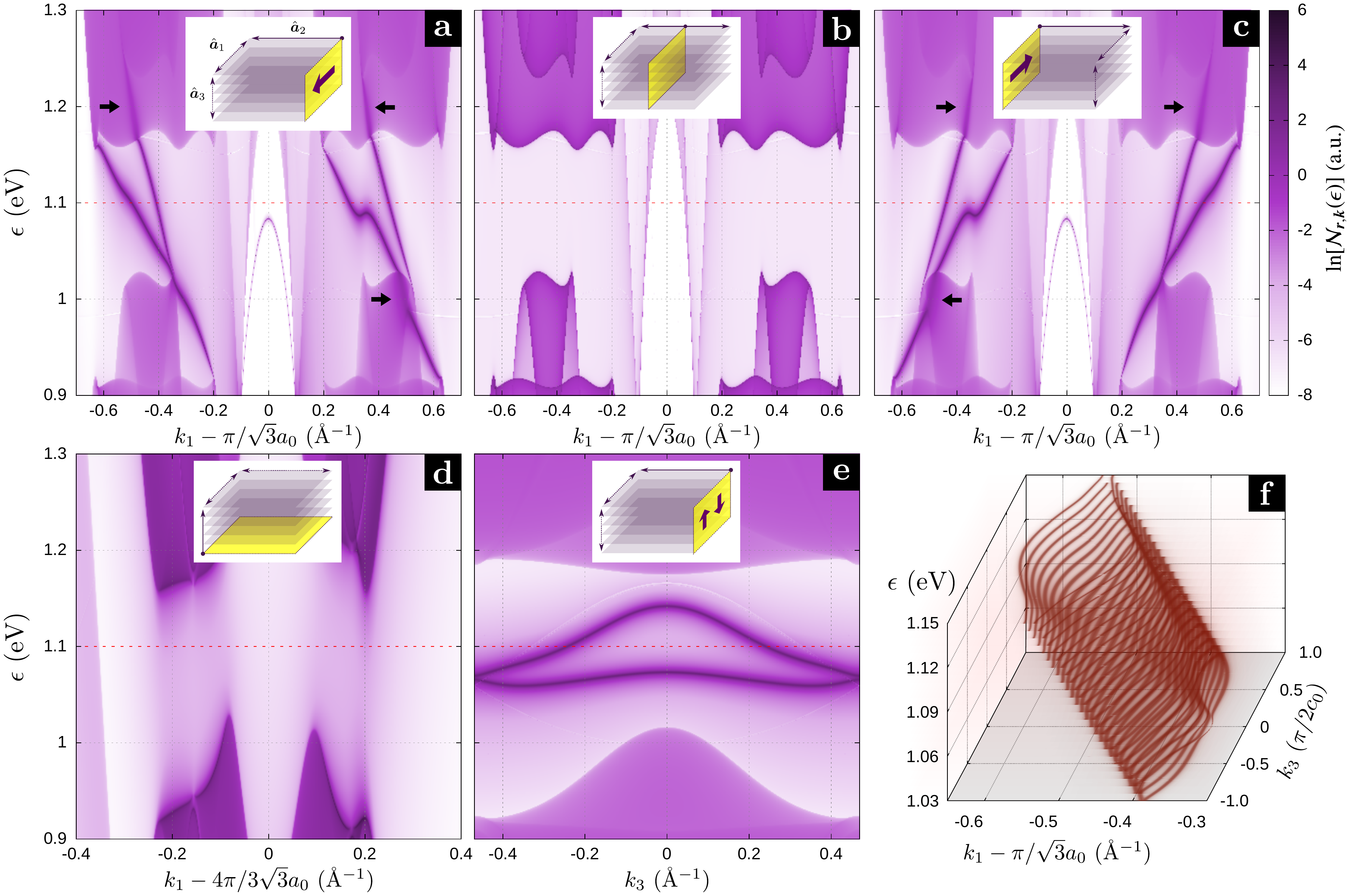}
\caption{Quasienergy and $\bm{k}$ resolved local density of states $\mathcal{N}_{\bm{r},\bm{k}}(\epsilon)$ in logarithmic scale for illuminated 
graphite. The LDoS is evaluated in a sample with $N$ unit cells along $\hbm{a}_2$ in (a)-(c) and (e) and along $\hbm{a}_3$ in (d). The insets schematically illustrate the regions in which the LDoS is being evaluated. Panels (a), (b) and (c) show the LDoS evaluated at $n_2 = 1$, $N/2$, and $N$, respectively, for $\bm{k} = k_1 \hbm{a}_1$; in (d) the LDoS is evaluated at $n_3 = 1$ for $\bm{k} = k_1 \hbm{a}_1$, while in (e) $n_2 = 1$ and $\bm{k} = (\pi/\sqrt{3}a_0+0.4 \ams^{-1}) \hbm{a}_1 + k_3 \hbm{a}_3$. We use an extremely large value for $N$, i.e. $N \sim 2^{25}$, such that the sample can be considered as semi-infinite in (a), (c)-(e), while in (b) the sample can be understood as infinite. In (f) we show the LDoS at $n_2 = 1$ as a function of energy and $\bm{k} = k_1 \hbm{a}_1 + k_3 \hbm{a}_3$. The laser parameters coincide with those in Fig.~\ref{fig:2} and the FZB at $\epsilon = 1.1$ eV is denoted by red dashed lines}.
\label{fig:3}
\end{figure*}

It is important to notice that for the gap at the FZC to be greater than that at the FZB one needs to be in the strong coupling regime. Therefore, as the laser intensity is smoothly increased from zero, the first visible feature, without the complications of heating and non-equilibrium effects present in the strong coupling regime,~\cite{sato2019} should be the gap at the FZB. Because of this, and also to keep within the validity range of our assumptions, we focus from now on in the FZB modes. Later on, when inspecting the bond currents in Sec.~\ref{sec:current}, we will set a stronger laser intensity where the FZC gap becomes clearly visible.

\section{Laser induced boundary states}\label{sec:bstates}

Having finished our program with illuminated bulk graphite, our next step is to check for laser-induced boundary states. To do it we introduce a boundary and inspect whether midgap states appear or not. We take one of the three directions of the lattice given by the primitive vectors of Eq.~(\ref{eq:primitive}) as finite, while keeping translational invariance along the other two. For example, we could define a `slab' geometry along the $\hbm{a}_1$ direction by taking a lattice with $\bm{R} = n_1 \bm{a}_1 + n_2 \bm{a}_2 + n_3 \bm{a}_3$, such that $1 \leq n_1 \leq N_1$, and $\{n_2,n_3\} \in \mathbb{Z}$. The problem then is that one should take a sufficiently large width as to prevent a considerable overlap between the expected 
boundary states, if these are present at the borders of the sample. This brings with it an important numerical effort since this geometry increases the dimension of the effective Hamiltonian to be diagonalized.

Perhaps a more convenient strategy to circumvent this issue is to refer to the time-averaged local density of states 
(LDoS) $\mathcal{N}_{\bm{r},\bm{k}}(\epsilon)$, which characterizes the weight of the $\bm{k}$-state at quasienergy $\epsilon$ on the site 
$\bm{r}$ along the broken direction. In the context of Floquet theory, this quantity can be written as~\cite{foatorres2005}
\begin{equation}
\mathcal{N}_{\bm{r},\bm{k}}(\epsilon) = -\frac{1}{\pi} \lim_{\eta \rightarrow 0^+} \tf{Im} \! \left[ \bra{\bm{r},0} \hat{\mathcal{G}}_{\tf{F},\bm{k}}(\epsilon+i\eta) \ket{\bm{r},0} \right],
\label{eq:ldos}
\end{equation}
with $\hat{\mathcal{G}}_{\tf{F},\bm{k}}$ the Floquet-Green operator associated with $\hat{\mathcal{H}}_{\tf{F},\bm{k}}$, i.e.,
\begin{equation}
\hat{\mathcal{G}}_{\tf{F},\bm{k}}(\epsilon) = \left[ \epsilon \hat{\mathcal{I}}_{\tf{F}} - \hat{\mathcal{H}}_{\tf{F},\bm{k}} \right]^{-1}.
\label{eq:green}
\end{equation}
The advantage of this method relies in that one still operates in the original dimension of the \textit{truncated} Floquet space, i.e., 
$\text{dim }\mathcal{F} = 4 \times (2n_r+1)$, where $n_r \geq 0$ denotes the highest taken value for the Fourier replica and we consider
the replicas going from $-n_r$ to $n_r$. The recursive Green's function method allows us to calculate the effective Hamiltonian of the unit 
cell placed at different positions within the sample,~\footnote{For example, if the slab is finite along $\hbm{a}_1$, we can evaluate 
$\mathcal{N}_{\bm{r},\bm{k}}$ for $\bm{r} = n_1 \bm{a}_1 + \bm{\delta}_i$, with $n_1 = 1,\dots,N_1$ and $\bm{\delta}_i$ the basis sites.} 
by including the self-energy corrections that account for the presence of all subsequent unit cells. This involves a decimation procedure which is 
explained in detail in Ref.~[\onlinecite{calvo2013}].

In Fig.~\ref{fig:3}, we show the illuminated graphite LDoS for different slab geometries as a function of the quasienergy $\epsilon$ and wavevector $\bm{k}$. Panels (a)--(c) and (e) refer to a sample which is finite along $\hbm{a}_2$, containing $N_2$ unit cells. In this case the corresponding Bravais lattice is rectangular, and given by primitive vectors $\bm{a}_1$ and $\bm{a}_3$. Therefore the primitive unit vectors of the reciprocal lattice coincide with those of the real lattice, and the wavevector can be written as $\bm{k} = k_1 \hbm{a}_1 + k_3 \hbm{a}_3$. We evaluate the LDoS at the positions $n_2 = 1$ in (a) and (e), $N_2/2$ in (b) and $N_2$ in (c), respectively. In panels (a)--(c), we take the wave vector as $\bm{k} = k_1 \hbm{a}_1$ and fixed $k_3 = 0$, while in panel (e) we use $\bm{k} = k_3 \hbm{a}_3$ and fixed $k_1 = \pi/\sqrt{3}a_0 + 0.4 \ams^{-1}$. The insets illustrate the regions where the LDoS is being evaluated: yellow rectangles denote the evaluation region and grey rectangles represent the graphene layers. In Fig.~\ref{fig:3}(d), we consider another geometry, where the sample is finite along $\hbm{a}_3$ and we evaluate the LDoS at the $n_3 = 1$ unit cell (see inset). In this case, the corresponding Bravais lattice is triangular, and we evaluate the LDoS for $\bm{k} = k_1 \hbm{a}_1$. As we use a huge value ($\sim 2^{25}$) for the amount of unit cells along the broken direction, the sample can be taken as semi-infinite in Figs.~\ref{fig:3}(a) and (c)-(e), while in Fig.~\ref{fig:3}(b) the sample is effectively infinite.

Figure~\ref{fig:3} shows the laser induced gap around $\epsilon = \hbar\Omega/2 = 1.1$ eV (see red dashed lines) and four states crossing the gap in (a) and (c), while these peaks in the LDoS disappear in (b). In panel (d), there is a clear gap induced by the laser at the FZB, and no peaks crossing this region can be observed. These are clear signals of the presence of laser induced boundary states, located at those surfaces perpendicular to the graphene layers [although not shown, figures similar to (a)--(c) are obtained for a finite sample along $\hbm{a}_1$]. The shape of the bands in Fig.~\ref{fig:3}(b) suggests that the laser produces two gaps centered around different quasienergies, which could be attributed to the four band structure observed in Fig.~\ref{fig:2}. The effective gapped region corresponds to the intersection between the two gaps, and outside this region these states may strongly mix with the bands [see black arrows in Figs.~\ref{fig:3}(a) and~\ref{fig:3}(c)]. From the slope of the trajectories defined by the LDoS peaks in Figs.~\ref{fig:3}(a) and~\ref{fig:3}(c), we can infer that these states propagate along the $\hbm{a}_1$ direction and with opposite velocities, depending on the border which is being evaluated. Specifically, the peaks shown in (a) can be attributed to states localized around the $n_2 = 1$ border that propagate along $-\hbm{a}_1$, while the peaks in (b) correspond to states localized around the $n_2 = N_2$ border which propagate along $+\hbm{a}_1$, see violet arrows in the inset schemes. In Fig.~\ref{fig:3}(e), we can observe the evolution of the localized states as we move the wavevector along the stacking direction, i.e. $\bm{k} = (\pi/\sqrt{3}a_0 + 0.4 \ams^{-1}) \hbm{a}_1 + k_3 \hbm{a}_3$, in the same spatial region as in Fig.~\ref{fig:3}(a), i.e. $n_2=1$. The peaks reveal some dispersion (non-negligible slope), meaning that the boundary states also propagate along the stacked layers. However, for a given border, these peaks stay in the middle of the gap without crossing it, and the slopes developed by them take both positive and negative values (a similar behavior occurs for $n_2 = N_2$). This means that the sign of the group velocity along the stacking direction is not restricted to the border in which the state is localized, so the two directions (say, positive and negative) may coexist in the same border (see violet arrows in the inset).

Notice that a similar behavior is obtained in monolayer graphene,~\cite{piskunow2014} where the circularly polarized laser induces chiral edge-states. By ``chiral'' is meant that the direction of propagation of the state depends on both the edge in which it is localized and the laser's handedness. In this sense, all the previous analysis indicates that in illuminated graphite there are also localized chiral states. We can continue this analogy and infer whether the observed localized states in graphite can be characterized by a topological invariant. This is presented in Appendix~\ref{sec:app_chern}, where we calculate the Chern number associated with the FZB for a simplified model of graphite that retains the leading hoppings $\gamma_0$, $\gamma_1$, and $\gamma_3$, and neglects all remaining (static) parameters. We are interested in the localized states generated by the mixing of the $n=0$ and $n=1$ replicas, so we truncate the Floquet space to these subspaces. Although higher-order mixings are also possible,~\cite{piskunow2015} the associated gaps decrease very fast for the considered small laser intensity, and these contributions can be neglected for the purpose of the present discussion. Under this approximation it is possible to derive analytic expressions for the eigenenergies of the bulk Hamiltonian of Eq.~(\ref{eq:bloch}), which allows us to identify the crossings between conduction and valence bands that belong to the $n=0$ and $n=1$ replicas, respectively. The main conclusion is that the contribution to the Chern number for a fixed value of $k_3$ is given by the number of bands that cross the FZB, multiplied by the sign $\tau$ of the polarization (thereby the chiral nature of these states). For the chosen frequency $\hbar\Omega = 2.2$ eV, this number results to be $4\tau$, in full agreement with the bulk-boundary correspondence, since the crossing bands are 1c--0, 1v--1, 2c--0 and 2v--1. Since we are computing only the contribution from the FZB gap to the topological invariant, it is implicitly assumed that the contributions from the bands below does not change. Considering all the contributions involves a more complex procedure as presented in Ref.~[\onlinecite{piskunow2015}] and is beyond our present scope.

So, what is new in this three-dimensional system? The first obvious difference with monolayer graphene is that now, rather than edge-states, what the LDoS peaks reveal are surface states located perpendicular to the planes defined by the graphene layers. To get an idea on how these surface states look, in Fig.~\ref{fig:3}(f) we evaluate the LDoS at $n_2 = 1$ for $\bm{k} = k_1 \hbm{a}_1 + k_3 \hbm{a}_3$ to picture out its shape in the \textit{two} directions where translational invariance holds. This was done by fixing the quasienergy in steps of 0.005 eV within the range 1.03 eV $\leq \epsilon \leq$ 1.15 eV. We use a transparency scale (arbitrary units) to visualize all $\bm{k}$-points where the LDoS takes a large value, such that the obtained curves define what can be thought of as the `skeleton' of the surface states. In fact, a close inspection for all energy steps when sweeping both $k_1$ and $k_3$ reveals two peaks which are separated each other, i.e. each peak defines an open trajectory. This suggests the presence of two surface states (in the shown region) which can be imagined as the natural dimensional extension of the chiral edge-states in graphene when adding an infinite number of layers. Another difference with monolayer graphene is that here the number of chiral states per value of $k_3$ is doubled, since now the gap comprises the crossing between four energy bands, due to the four basis sites in the unit cell. This, however, may change depending on the value of the chosen frequency. When $\hbar\Omega/2 \lesssim 0.25 \gamma_0$ and $k_3 \sim 0$, it may happen that the bands that cross at the FZB are only 1c--0 and 1v--1, so the expected Chern number is in this case $2\tau$ (see Appendix~\ref{sec:app_chern}).

\begin{figure}[t]
\includegraphics[width=\columnwidth]{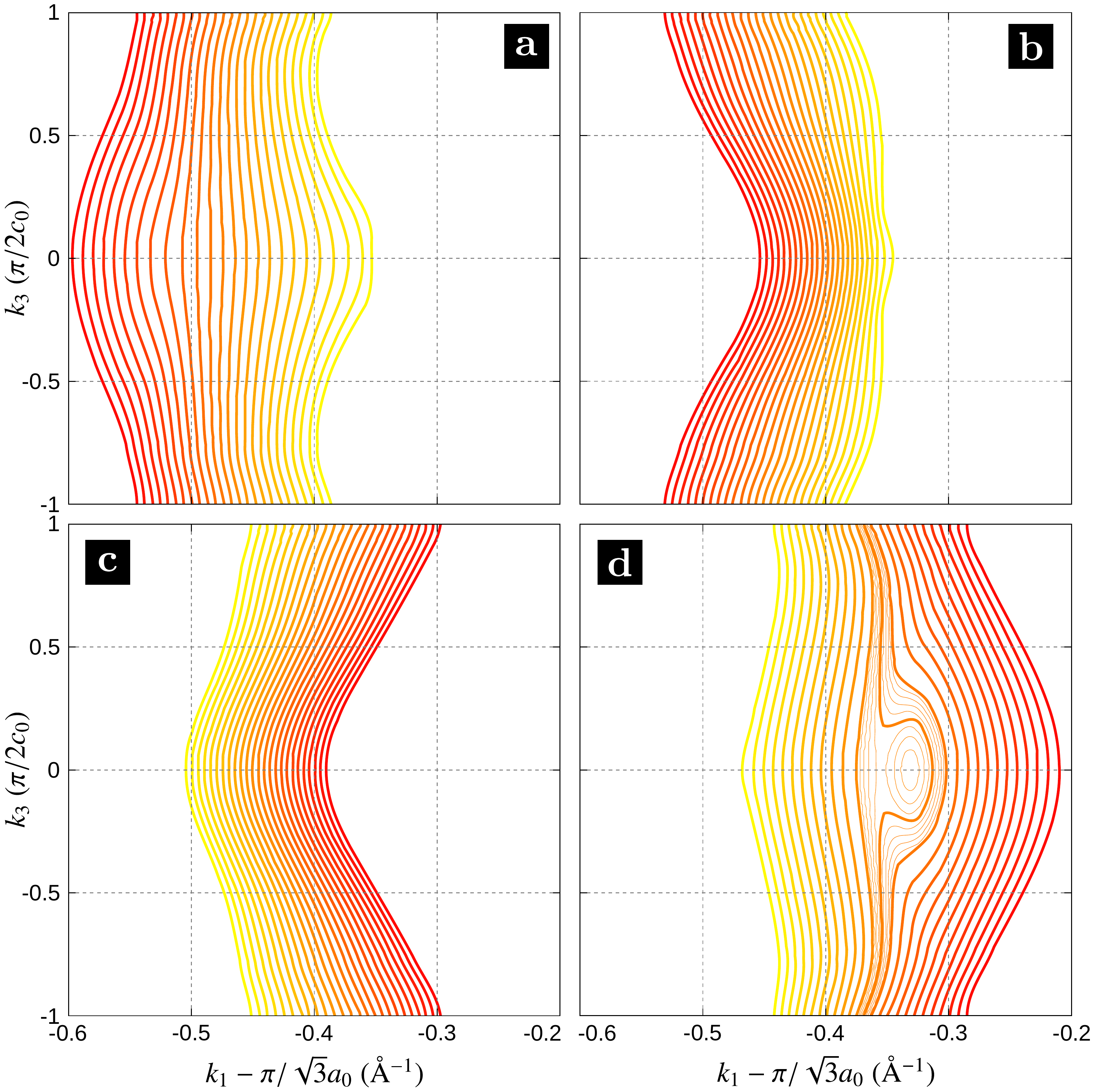}
\caption{Laser induced surface states extracted from the LDoS. (a) and (b) show the two peaks of Fig.~\ref{fig:3}(f), respectively, corresponding to the LDoS at $n_2=1$. Similarly, (c) and (d) show the LDoS peaks when evaluated at $n_2 = N$. The used colorscale goes from red ($\epsilon = 1.15$ eV) to yellow ($\epsilon = 1.03$ eV).}
\label{fig:4}
\end{figure}

In Fig.~\ref{fig:4} we extract the maxima of the peaks of Fig.~\ref{fig:3}(f) and separate them in panels (a) and (b) to appreciate the surface states individually. The same is done in (c) and (d) for the LDoS evaluated at $n_2 = N_2$. The lines thus correspond to those boundary states that form the surface state for a fixed energy. We use the same energies as in Fig.~\ref{fig:3}(f) but these are distinguished through a colorscale ranging from red ($\epsilon = 1.15$ eV) to yellow ($\epsilon = 1.03$ eV). The shown plots thus resemble maps of equipotential lines (quasienergies) associated to the surface states. From the colorscale it is possible then to infer the group velocity of these states. Since 
\begin{equation}
\bm{v}_{\tf{g}}(\bm{k}) = \frac{1}{\hbar}\nabla_{\bm{k}} \epsilon_{\bm{k}},
\end{equation}
the group velocity points from yellow to red and perpendicular to the equipotential lines. From the plots it is easy to see that, in almost all cases, $\bm{v}_\tf{g}$ points along the horizontal axis, i.e. $\hbm{a}_1$. In Fig.~\ref{fig:4}(d) there is, however, a particular region where a local minimum develops,~\footnote{We increased the resolution in steps of $\epsilon = 0.001$ eV within the range 1.085 eV $\leq \epsilon \leq$ 1.095 eV to identify this local minimum.} and the $\hbm{a}_3$ component of the group velocity dominates over the $\hbm{a}_1$ component at least locally. In any case, the mirror symmetry around $k_3 =0$ implies that
\begin{equation}
\bm{v}_{\tf{g}}(k_1,k_3) \cdot \hbm{a}_3 = -\bm{v}_{\tf{g}}(k_1,-k_3) \cdot \hbm{a}_3, 
\end{equation}
meaning that for a given Fermi energy within the gapped region the overall velocity points along $\hbm{a}_1$ only. With this in mind, we can again conclude that these states are chiral, since the group velocity points towards opposite directions regarding to which border the surface state belongs.

All the above findings therefore enforce the idea that the physics behind illumination on graphite is, to some extent, similar to that of monolayer (or bilayer) graphene. The additional dimension present in this case contributes with a weak component of the group velocity along the new direction, which averages to zero when populating the system to the FZB. This is possibly due to the large separation between the stacked layers ($c_0$), as compared to the first-neighbor distance ($a_0$). The obtained surface states are rather continuations of graphene's edge-states in the stacking direction so in this sense one could say that these move through the boundary of the sample in an orderly manner. To which extent this is true is a question whose answer requires the evaluation of the LDoS along the boundary when both the $\bm{a}_2$ and $\bm{a}_3$ directions are finite. This obviously difficults the calculation of the LDoS as the effective dimension over which one operates is now dim $\mathcal{F} = 4 N_3 \times (2n_r+1)$, where $N_3$ is the number of unit cells along $\hbm{a}_3$. For small samples ($N_3 \sim 10$) this can be done in the same way we did before (i.e. an exact calculation), but for larger samples the previous strategy becomes very demanding (computationally speaking) and we employ an approximation scheme based on a decomposition into normal modes similar to that used in Refs.~[\onlinecite{rocha2010}] and~[\onlinecite{calvo2018}]. Although in graphite this decomposition scheme is not exact due to the next-nearest-neighbor couplings $\gamma_2$ and $\gamma_5$, deviations from the exact result can be considered as a small perturbation acting only on $n_3 = 1$ and $n_3 = N_3$, which can be neglected in large samples. This we explain in further details in App.~\ref{sec:app_modes}.

\begin{figure*}[th]
\centering
\includegraphics[width=\textwidth]{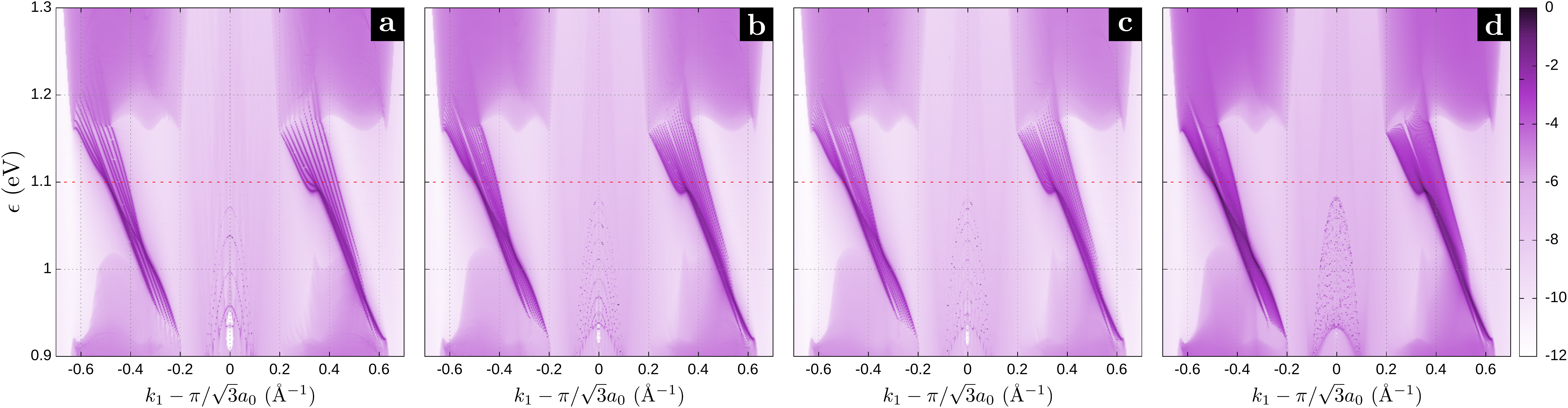}
\caption{LDoS evaluated at $n_2=1$ for broken $\hbm{a}_2$ and $\hbm{a}_3$ directions. (a) and (b) are $k$ vs $\epsilon$ LDoS maps for $N_3 = 5$ and 10, respectively, obtained from an exact calculation. (c) and (d) show the approximated LDoS through the normal mode decomposition for $N_3 = 10$ and 100, respectively. In all plots we normalized the densities to its maximum value and used a logarithmic scale. The red dashed lines at $\epsilon = 1.1$ eV denote the FZB.}
\label{fig:5}
\end{figure*}

In Fig.~\ref{fig:5}, we show the LDoS evaluated at $n_2 = 1$ for different sample sizes, given by the number $N_3$ of unit cells along $\hbm{a}_3$. Panels (a)--(d) are the maps in the same $k_1$ region as in Fig.~\ref{fig:3}(a). In (a) and (b), we used the standard decimation procedure as in all previous calculations, while in (c) and (d), we used the normal mode decomposition explained in Appendix~\ref{sec:app_modes}. A comparison between Figs.~\ref{fig:5}(b) and~\ref{fig:5}(c) for $N_3 = 10$ shows that the used decomposition, though not exact, yields an accurate LDoS even in relatively small samples. 

As expected, we can see that the peaks of Fig.~\ref{fig:3} are also present in this case, maintaining the same chiral behavior as before. This is somewhat obvious when regarding the LDoS as decomposed by normal modes along $\hbm{a}_3$. Since this decomposition takes discrete values of $k_3$, c.f. Eq.~(\ref{eq:cuts}), the LDoS for a fixed $k_3$ is similar to that of Fig.~\ref{fig:3}(a), and the final LDoS is given as the sum of all mode contributions. For the considered region in the maps, then, the number of chiral edge-states crossing the gap simply goes as $4N_3$, as anticipated by the total Chern number of App.~\ref{sec:app_chern}. This is easy to see when $N_3$ is small, as it happens in Figs.~\ref{fig:5}(a)--\ref{fig:5}(c). For $N_3 = 100$, however, such a counting is no longer possible in Fig.~\ref{fig:5}(d) even if we would be able to increase the map resolution indefinitely. The reason for this is a rather subtle effect we did not comment so far. All LDoS peaks we have shown have, in fact, a finite width, which is independent of the chosen regularization energy $\eta$ of Eq.~(\ref{eq:ldos}).~\footnote{In all LDoS calculations, we use $\eta = 1 \times 10^{-5}$ eV, as this reaches the thermodynamic limit for the number $N \sim 2^{25}$ of considered unit cells.} To understand the origin of this width, notice that the localized states around $\epsilon = \hbar \Omega/2$ are produced by the coupling between $n=0$ and $n=1$ replicas. However, other extended states belonging to other replicas may be present within the gapped region. Strictly speaking, there is no real gap in the FZB where the localized states develop. However, we refer to the opening of $n=0$ and $n=1$ bands as a ``gap'' since the contributions coming from other replicas to the time-averaged LDoS are quite small. In other words, only when the replicas $n=0$ and $n=1$ are considered, the band opening at the FZB is a real gap. The observed width in the LDoS peaks then signals a small mixing term between the localized states (formed as a superposition of the $n=0$ and $n=1$ replicas) and extended states from other Floquet replicas (in this case the  main contribution comes from $n=-1$ and $n=2$). This implies that the localized states decay into the bulk upon absorption or emission of photons, in a characteristic time proportional to the inverse of the energy width of the peaks. Therefore, when the number of localized states is small, the mean level spacing is larger than their widths, and the system ``recognizes'' its finite size along $\hbm{a}_3$. When increasing $N_3$, at some point the level spacing becomes comparable to the energy width, and the system is no longer able to discern its finite size, so it behaves as a bulk in the staking direction. This originates the formation of localized states bands of Fig.~\ref{fig:5}(d), which may well be taken as surface states even in this limit of relatively small $N_3$.

\section{Laser induced probability currents in finite systems}\label{sec:current}

Another interesting effect that we would like to address is the fact that chiral states, by having a well-defined direction of propagation, are able to transport a probability current along the sample. This was shown in the context of illuminated monolayer graphene, where the laser-induced probability current appears either along the borders of the sample~\cite{dahlhaus2015} or surrounds different types of defects like vacancies and adatoms.~\cite{lovey2016} Interestingly, such edge-states and their associated currents are able to be accessed by measuring the magnetic field they produce.~\cite{dahlhaus2015} In graphite, therefore, similar effects can be naturally expected. To illustrate this, we consider a finite graphite sample consisting in a few hexagonal layers along the stacking direction. According to the discussion in Sec.~\ref{sec:floquet}, the quantity of interest, rather than the site current $J(\bm{r})$, is the bond current $J(\bm{r},\bm{r}')$ given by Eq.~(\ref{eq:current}).~\cite{todorov2002} The carbon bonds where this current is non-zero are thus given by those sites coupled by the Floquet Hamiltonian.

\begin{figure*}[th]
\centering
\includegraphics[width=\textwidth]{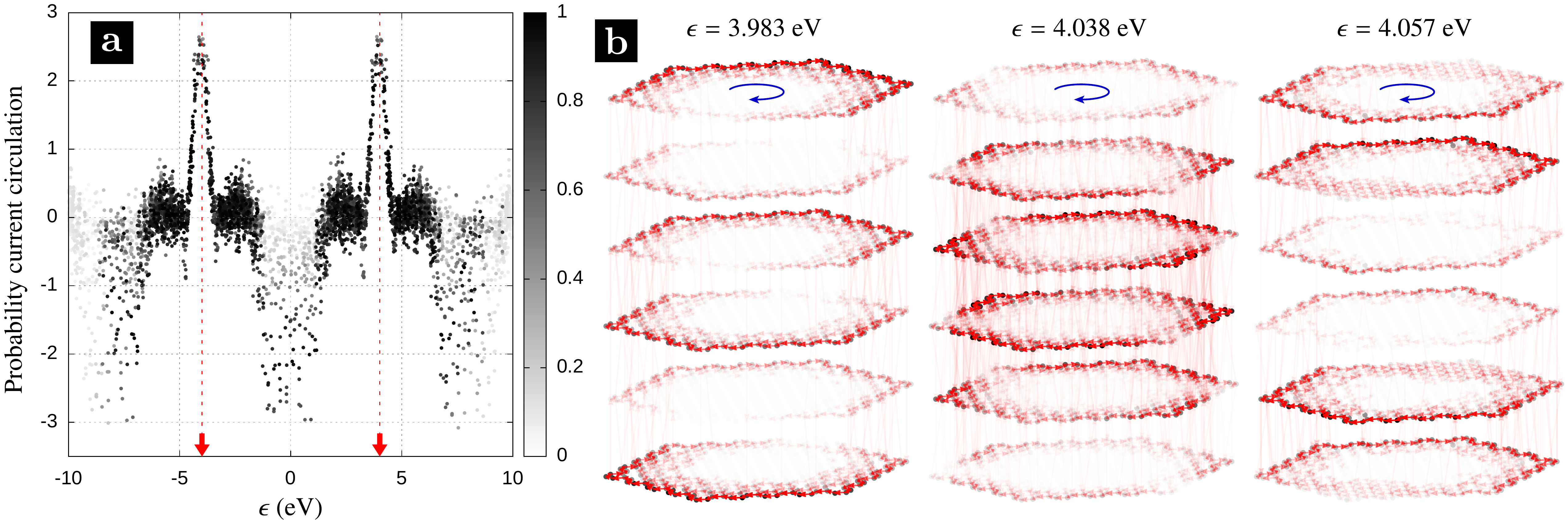}
\caption{(a) Circulation of the time-averaged probability current as a function of the quasienergy in arbitrary units. The gray scale emphasizes the mixing of the $n=0$ replica (see text). The red dashed lines at $\epsilon = \pm 4$ eV denote the FZB. (b) Examples of laser induced probability currents in few-layer graphite. The chosen geometry for the layers is hexagonal and we used $N_3 = 3$, i.e. six graphene layers. The position of the carbon atoms is represented by dots and we use a gray scale to denote the time-averaged probability density of Eq.~(\ref{eq:rho}). The corresponding probability bond currents (red arrows) are plotted in a transparency scale according to their magnitude. Blue arrows indicate the overall direction of the bond currents. The laser parameters were changed to $\hbar\Omega = 8$ eV and $\zeta_0 = 0.5 \ams^{-1} \times 1.42 \ams = 0.71$.}
\label{fig:6}
\end{figure*}

\begin{figure}[th]
\centering
\includegraphics[width=\columnwidth]{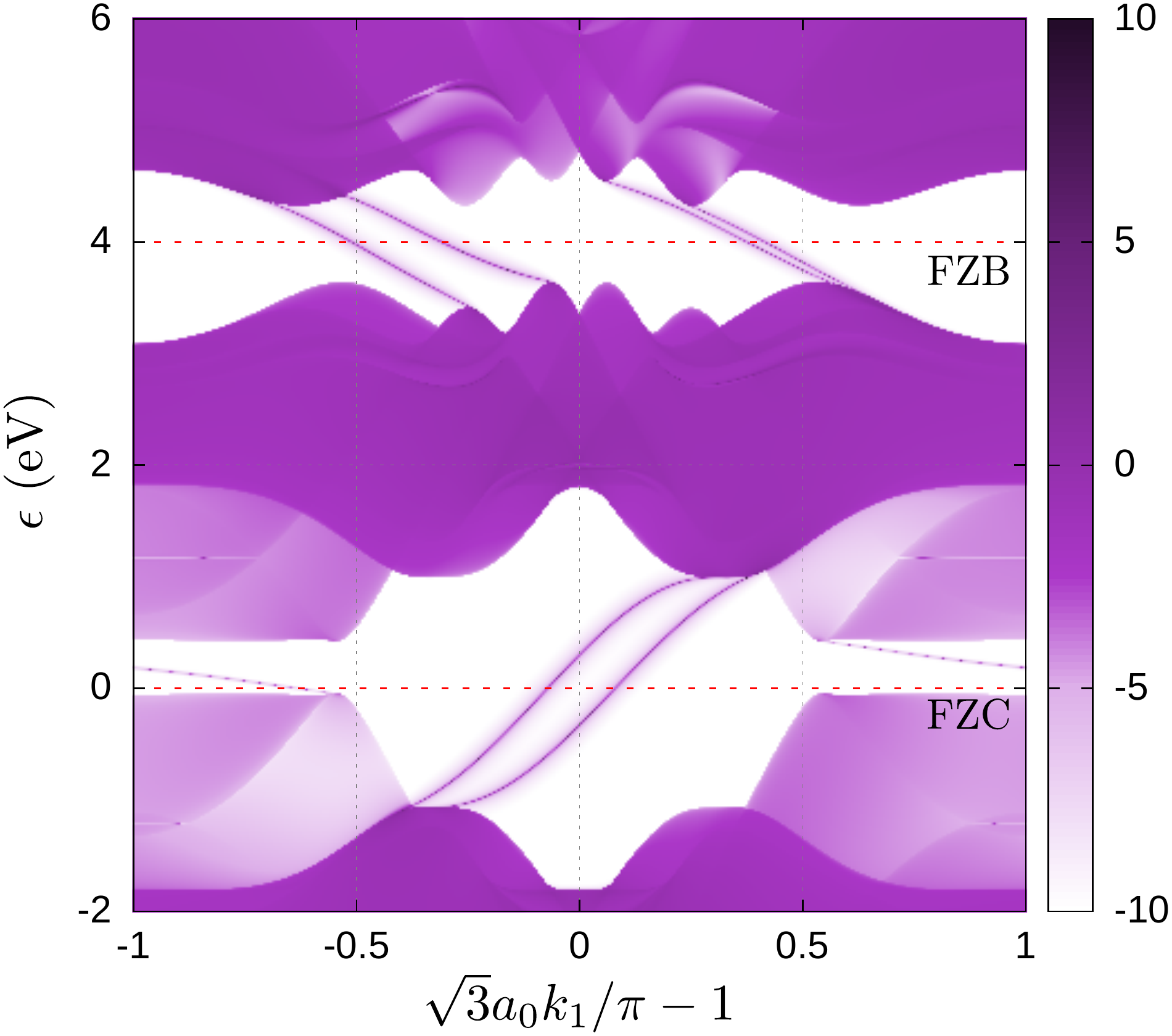}
\caption{LDoS (in logarithmic scale) evaluated at $n_2 = 1$ for broken $\hbm{a}_2$ direction as in Fig.~\ref{fig:3}a. The used laser parameters coincide with those of Fig.~\ref{fig:6}. Red dashed lines at $\epsilon = 0$ and $4$ eV denote the FZC and FZB, respectively.}
\label{fig:7}
\end{figure}

In order to identify the role of the laser illumination on the chirality of these currents, we also calculate the circulation of the bond currents through the lateral borders of the sample. This can be computed as the following discrete version of the line integral of the bond currents:
\begin{equation}
C_\alpha = \sum_{\bm{r},\bm{r}' \in S} J_\alpha(\bm{r},\bm{r}'),
\end{equation}
where $\alpha$ labels the Floquet state and the sum runs over all sites belonging to the border of the layers. In Fig.~\ref{fig:6}(a) we show the obtained circulation of the probability current for the hexagonal sample shown in Fig.~\ref{fig:6}(b). The dots in the plot are the obtained quasienergies from the eigenvalue equation, and we used a grayscale to indicate the weight of the Floquet eigenstate $\ket{\alpha}$ on the $n=0$ replica, but with a minor change with respect to Eq.~(\ref{eq:weight}), i.e.
\begin{equation}
\bar{w}'_\alpha = 1 - 2 \left| \bar{w}_\alpha - \frac{1}{2} \right|, \text{ where } \bar{w}_\alpha = \sum_{\bm{r}} | \! \braket{\bm{r},0|\alpha} \! |^2.
\end{equation}
The idea behind this modification is to highlight the superposition of the $n=0$ replica with the remaining ones: when $\bar{w}_\alpha = 1$, the Floquet state has full weight on the $n=0$ replica, so there is no mixing with higher-order replicas ($\bar{w}'_\alpha = 0$), and when $\bar{w}_\alpha = 0$  the state has no weight on $n = 0$ and so again there is no mixing. The maximum value $\bar{w}'_\alpha = 1$ is reached when $\bar{w}_\alpha = 1/2$, meaning that the probability to find the system in the $n=0$ replica is equal to that of finding it in all other replicas. Roughly speaking, $\bar{w}'_\alpha$ serves to infer where photon emission/absorption processes are more likely to occur.
For the calculations, we used 366 carbon atoms per layer, so the total dimension of the truncated Floquet space is dim $\mathcal{F} = 366 \times 6 \times 5$, where the six corresponds to the number of layers and we considered five Floquet replicas, i.e. $n=-2,\dots,2$. For this example then, diagonalization of the Floquet Hamiltonian is a problem that can be treated exactly. However, as we discussed before in the context of the LDoS, for larger samples such a calculation may become seriously hard and one should move to the normal mode decomposition of Appendix~\ref{sec:app_modes}. We here took $\hbar\Omega = 8$ eV and $\zeta_0 = 0.71$ for the laser's parameters. Though these parameters may exceed standard values, we use them as to illustrate the effect in a relatively small sample. The same effects would be obtained for smaller parameters when used in larger samples, specially the size of the hexagonal layers, were a mode decomposition is not available.

In order to support the obtained circulation of the bond currents we show, in Fig.~\ref{fig:7}, the LDoS for the same geometry as that used in Fig.~\ref{fig:3}(a). This allows us to identify the boundary states appearing at the FZB and FZC gaps in this regime of laser parameters.

The main feature of Fig.~\ref{fig:6}(a) are the peaks of $C_\alpha$ in the vicinity of the FZB, defined at $\epsilon = \pm \hbar\Omega/2$ (red dashed lines). In \textit{all} states within this region, the probability density circulates through the boundaries of the hexagonal layers, with a given handedness. Obviously, if we change the sign of the circularly polarized waves, the direction of the bond currents is inverted, and with it the sign of the circulation. In addition, the mixing of the $n=0$ replica in the peaks is large (i.e., $\bar{w}'_\alpha \sim 1$), which indicates 
a correlation between circulation and photon emission/absorption processes. In other words, the illuminated electrons are more likely to circulate in the energy regions where the interaction with the laser field becomes relevant. This also happens around the FZC, defined at $\epsilon = 0$. Although here the band crossings are more complicated than in the FZB, we can appreciate a negative circulation, though not all states are participating in this peak. In fact, we can identify some states with small (or even positive) circulation, which accordingly are weakly mixed. This can be attributed to the boundary states appearing in the FZC gap of Fig.~\ref{fig:7}, where we can see two states with positive group velocity and a single state with negative group velocity. Of course, the comparison between Figs.~\ref{fig:6}(a) and \ref{fig:7} can only be taken as qualitative, since for the LDoS we used a semi-infinite sample along the $\hbm{a}_2$ direction while for the bond currents we used a finite system.

In Fig.~\ref{fig:6}(b), we show the bond currents and probability densities for three Floquet states whose eigenenergies lie close to $\hbar\Omega/2$. The bond currents $J(\bm{r},\bm{r}')$ are shown in red arrows that go from $\bm{r}'$ to $\bm{r}$, and we use a transparency scale to indicate its relative magnitude to the maximum current. For each carbon atom, we also calculated the time-averaged probability density $\rho(\bm{r})$ given in Eq.~(\ref{eq:rho}) and is shown through a gray scale. The resulting Floquet states around this energy region are clearly localized at the boundaries of the hexagonal layers. As Eq.~(\ref{eq:current}) suggests, the bond currents are expected to be nonzero in those sites where $\rho(\bm{r})$ is appreciable, so they are also confined to the boundaries of the layers.

The bond currents' features discussed in this section are clear fingerprints of the chiral nature of the laser induced localized states. Interestingly, some differences appear when comparing these states with those found in illuminated monolayers. In fact, the magnitude of the bond currents in Fig.~\ref{fig:6}(b) is not constant along the full border of the hexagonal layers, but it rather alternates between successive layers. This is provided by some small, but non negligible, bond currents pointing along the stacking direction. Although this effect does not break the chirality of the localized states, the probability current displays nontrivial patterns due to the interlayer hopping amplitudes.

\section{Summary and final remarks}

To sum up, illumination by a circularly polarized laser on graphite generates boundary states. These boundary states turn out to be chiral, may form bands bridging the gap, and bear similarities and differences with those found in graphene. In the limit of large samples we show that a normal mode decomposition is applicable along the vertical direction. This provides a useful tool to reduce the 3D system onto a set of decoupled 2D subsystems where the $z$ component of the wavevector enters as a fixed parameter. Under this decomposition scheme we were able to calculate the corresponding Chern number, which can be linked to the number of bands that intersect at the symmetry point $\epsilon = \hbar\Omega/2$. We highlight, however, two interesting features which we attribute to the extra dimension of the sample. First, we observe a smooth transition in the local density of states that goes from separable peaks (bundles) to the formation of bands of surface states, which evidence the three-dimensional nature of the sample even for relatively small $N_3$ values. This is attributed to a photon assisted decay of the localized states into extended states that belong to higher-order replicas. Second, the calculated probability currents may display intrincate patterns due to the small component along the stacking direction.

Regarding other possible stacking orders for graphite it should be noticed that, in principle, each crystal structure could present a topological structure of its own. However, given the hierarchical layered structure of graphite, we expect that in this case the main features observed for AB stacking should be kept. Notwithstanding, this is beyond the scope of our study, which remains non-exhaustive in this respect, motivating further investigations on illuminated multilayered systems.

We hope that the obtained results may stimulate further experimental research in strong light-matter interaction in graphite and related systems.

\vspace{0.5cm}

\noindent
\textit{Acknowlegdments.--} This work was supported by Consejo Nacional de Investigaciones Cient\'ificas y T\'ecnicas (CONICET), Secretar\'ia de 
Ciencia y Tecnolog\'ia -- Universidad Nacional de C\'ordoba (SECYT--UNC). HLC is member of CONICET. LEFFT acknowledges funding from FondeCyT (Chile) under project number 1170917 and 1170921, and by the EU Horizon 2020 research and innovation program under the Marie-Sklodowska-Curie Grant Agreement No. 873028 (HYDROTRONICS Project). JEBV acknowledges funding from PAIP Facultad de Química UNAM (grant 5000-9173).

\appendix
\section{Chern number calculation} \label{sec:app_chern}

In this section we sketch the calculation of the Chern number associated with the light induced band-gap openings around the crossing region $\epsilon = \hbar\Omega/2$, i.e., the Floquet zone boundary (FZB). To such end, we derive analytic expressions for the energies of those bands crossing at the FZB, under the mode decomposition scheme presented in App.~\ref{sec:app_modes}. This implies that in the present model we consider the most relevant hopping terms $\gamma_0$, $\gamma_1$ and $\gamma_3$, and neglect all remaining ones in the bulk Hamiltonian of Eq.~(\ref{eq:bloch}). The contributions $c_p$ to the Chern numbers are given by each band crossing $p$ taking place at the FZB, when the laser is turned off, and can be obtained by reducing the Hamiltonian to those bands participating in the crossing. This yields a $2 \times 2$ effective Hamiltonian of the form
\begin{equation}
\hat{\mathcal{H}}_{p,\tf{eff}} = \bm{h}_p \cdot \hbm{\sigma}, \label{eq:chern}
\end{equation}
with $\hbm{\sigma}$ the vector of Pauli matrices and $\bm{h}_p$ the associated vector to the $p$-crossing. The corresponding expression for $c_p$ is the following~\cite{hasan2010}
\begin{equation}
c_p = \frac{1}{4\pi} \int \tf{d}^2\bm{k} \; \hbm{h}_p \cdot \left( \partial_{k_x} \hbm{h}_p \times \partial_{k_y} \hbm{h}_p \right), 
\end{equation}
where the integral is taken over the first Brillouin zone for $k_3$ fixed as in Eq.~(\ref{eq:cuts}) and $\hbm{h}_p$ is the unit vector associated with $\bm{h}_p$. In order to obtain $\bm{h}_p$, we start with the bulk Floquet Hamiltonian of Eq.~(\ref{eq:bloch-floquet}), and truncate the Floquet space to replicas $n=0$ and $n=1$. This can be computed as the following matrix:
\begin{equation}
\bm{H}_\tf{F} = \left(
\begin{array}{cc}
\bm{H}^{(0)}-w	& \bm{H}^{(1)} 		\\
\bm{H}^{(-1)}	& \bm{H}^{(0)}+w
\end{array} \right),
\end{equation}
where we shifted the energy origin to $w=\hbar\Omega/2$ so that the crossings we are interested in are placed at $\epsilon = 0$. The structure of these block matrices obey the form given in Eq.~(\ref{eq:bloch}), i.e.
\begin{equation}
\bm{H}^{(n)} = \left(
\begin{array}{cccc}
0					& \gamma_{12}^{(n)} & \gamma_{13}^{(n)} & 0					\\
\gamma_{21}^{(n)}	& 0 				& 0 				& \gamma_{24}^{(n)}	\\
\gamma_{31}^{(n)}	& 0 				& 0 				& \gamma_{34}^{(n)}	\\
0					& \gamma_{42}^{(n)} & \gamma_{43}^{(n)} & 0
\end{array} \right).
\end{equation}
For the calculation of the hopping terms, we assume $\zeta_0 \ll 1$ so the phase introduced by the vector potential in Eq.~(\ref{eq:hopp1}) can be linearized as
\begin{equation}
e^{i2\pi(\bm{r}-\bm{r}') \cdot \bm{A}(t)/\Phi_0} \simeq 1+i \zeta_0 \cos(\Omega t - \tau \phi_{\bm{r},\bm{r}'}),
\end{equation}
where $\tau = \pm 1$ denotes the laser's handedness. Following Eq.~(\ref{eq:hopp1}), we notice that in all cases we have $|\bm{r}-\bm{r}'| \sin\theta_{\bm{r},\bm{r}'} = a_0$, and hence $\zeta_{\bm{r},\bm{r}'}=\zeta_0$. Recalling that in the construction of the Floquet Hamiltonian we multiply these terms by $\exp(i n \Omega t)$ and take the time-integral over one period, this yields for the above equation:
\begin{equation}
\delta_{n,0} + i \frac{\zeta_0}{2} e^{i n\tau\phi_{\bm{r},\bm{r}'}} (\delta_{n,-1}+\delta_{n,1}).
\end{equation}
Therefore, the hopping terms can be specified by:
\begin{align*}
\gamma_{21}^{(0)}	&=	\gamma_0 \left(1+e^{-i\bm{k}\cdot\bm{a}_1} + e^{-i\bm{k}\cdot\bm{a}_2} \right), \\
\gamma_{21}^{(1)}	&=	i\frac{\zeta_0 \gamma_0}{2} \left(e^{+i \tau \frac{1}{2}\pi} + e^{-i(\bm{k}\cdot\bm{a}_1+ \tau \frac{5}{6}\pi)}
						+e^{-i(\bm{k}\cdot\bm{a}_2 + \tau \frac{1}{6}\pi)}\right), \\
\gamma_{12}^{(1)}	&=	i\frac{\zeta_0 \gamma_0}{2} \left(e^{-i \tau \frac{1}{2}\pi} + e^{+i(\bm{k}\cdot\bm{a}_1+ \tau \frac{1}{6}\pi)}
						+e^{+i(\bm{k}\cdot\bm{a}_2 + \tau \frac{5}{6}\pi)}\right),
\end{align*}
together with
\begin{align*}
\gamma_{31}^{(n)}	&= \gamma_1 \left(1+e^{-i\bm{k}\cdot\bm{a}_3}\right)\delta_{n,0},\\
\gamma_{43}^{(n)}	&= \gamma_{12}^{(n)}e^{-i\bm{k}\cdot\bm{a}_1},\\
\gamma_{42}^{(n)}	&= \frac{\gamma_3}{\gamma_0}\gamma_{21}^{(n)}e^{+i\bm{k}\cdot\bm{a}_2}\left(1+e^{-i\bm{k}\cdot\bm{a}_3}\right),
\end{align*}
and the general rule $\gamma_{ij}^{(n)} = [\gamma_{ji}^{(-n)}]^\ast$. With all these terms specified, we now construct the above Floquet Hamiltonian, and diagonalize the blocks $\bm{H}^{(0)}$. This gives the following eigenenergies:
\begin{equation*}
\epsilon_{1,\tf{c}} = \sqrt{\frac{\alpha}{2}-\sqrt{\frac{\alpha^2}{4}-\beta}}, \quad \epsilon_{2,\tf{c}} = \sqrt{\frac{\alpha}{2}+\sqrt{\frac{\alpha^2}{4}-\beta}},
\end{equation*}
for the conduction bands, while for the valence bands we have $\epsilon_{p,\tf{v}} = -\epsilon_{p,\tf{c}}$ for $p={1,2}$, as in this model the e-h symmetry is preserved when $\gamma_2$, $\gamma_4$ and $\gamma_5$ are neglected. The terms in the above expressions are given by:
\begin{align*}
\alpha &= |\gamma_{21}|^2+|\gamma_{31}|^2+|\gamma_{42}|^2+|\gamma_{43}|^2, \\
\beta  &= |\gamma_{21}|^2|\gamma_{43}|^2+|\gamma_{31}|^2|\gamma_{42}|^2-2 \tf{Re}\left(\gamma_{13}\gamma_{34}\gamma_{42}\gamma_{21}\right),
\end{align*}
where we simplified the notation by taking $\gamma^{(0)} \rightarrow \gamma$, i.e., all hoppings in $\alpha$ and $\beta$ correspond to the zeroth Fourier component. When including the $w$-term in these bands, we obtain the following two crossings:
\begin{equation}
\epsilon_{1,\tf{c}}-w = \epsilon_{1,\tf{v}}+w, \quad \tf{and} \quad \epsilon_{2,\tf{c}}-w = \epsilon_{2,\tf{v}}+w. \label{eq:cond-w}
\end{equation}
So we have that the bands that participate in the crossings are the conduction bands associated to the $n=0$ replica and the valence bands for the $n=1$ replica. The above mentioned e-h symmetry implies that the crossing conditions are simply given by $\epsilon_{p,\tf{c}}=w$, where $p={1,2}$ now labels each band crossing. 

\begin{figure}[t]
\centering
\includegraphics[width=0.9\columnwidth]{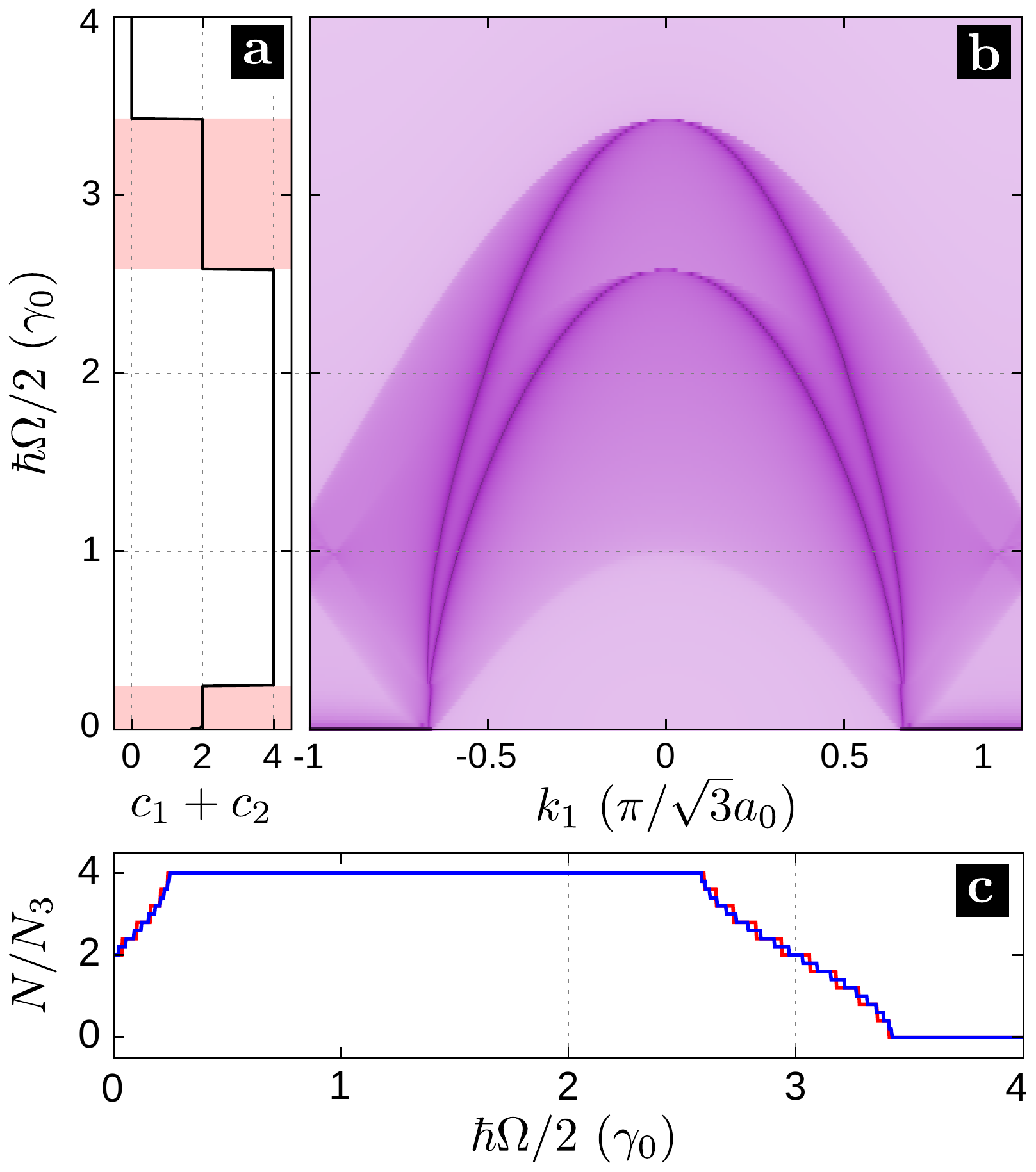}
\caption{Localized states and Chern numbers as a function of the driving frequency $\Omega$. (a) Contributions to the Chern number from the crossings $p=1$ and $p=2$. The red shaded areas denote transition regions where the Chern number may vary depending on the value of $k_3$. The laser intensity is $\zeta_0 = 0.01$ and we took $k_3 = 0$. (b) LDoS for broken $\bm{a}_2$ direction and evaluated at $n_2=1$ for $\epsilon = \hbar\Omega/2$. (c) Number of bands crossing at the FZB, divided by $N_3$, for $N_3=5$ (red) and $N_3=10$ (blue), in the limit $\zeta_0 = 0$.}
\label{fig:app}
\end{figure}

The following step is to reduce the Floquet Hamiltonian to the found crossings. What we obtain then is the effective Hamiltonian as
\begin{equation}
\hat{\mathcal{H}}_{p,\tf{eff}} = \left(
\begin{array}{cc}
\epsilon_{p,\tf{c}}-w	& \gamma_{\tf{c},\tf{v}}^{(p,1)} \\
\gamma_{\tf{v},\tf{c}}^{(p,-1)}	& \epsilon_{p,\tf{v}}+w	
\end{array} \right),
\end{equation}
where $\gamma_{\tf{v},\tf{c}}^{(p,-1)}$ and $\gamma_{\tf{c},\tf{v}}^{(p,1)}$ are obtained after applying the transformation matrix $\bm{U}$ that diagonalizes $\bm{H}^{(0)}$ on the coupling matrices $\bm{H}^{(\pm 1)}$ between the replicas, i.e. 
\begin{align}
\gamma_{\tf{v},\tf{c}}^{(p,-1)} &= \bra{p,\tf{v},1} \bm{U}^\dag \bm{H}^{(-1)} \bm{U} \ket{p,\tf{c},0}, \\
\gamma_{\tf{c},\tf{v}}^{(p,+1)} &= \bra{p,\tf{c},0} \bm{U}^\dag \bm{H}^{(+1)} \bm{U} \ket{p,\tf{v},1}.
\end{align}
In this way, we can identify the components of the $\bm{h}_p$ vector multiplying the Pauli matrices in Eq.~(\ref{eq:chern}) as
\begin{equation}
\bm{h}_p = \left(\tf{Re} [\gamma_{\tf{v},\tf{c}}^{(p,-1)}], \tf{Im} [\gamma_{\tf{v},\tf{c}}^{(p,-1)}], \epsilon_{p,\tf{c}}-w \right).
\end{equation}
What follows in the calculation of $c_p$ are the $k_x$ and $k_y$ derivatives, together with the integration over the first Brillouin zone. We carried out this numerically and obtained the Chern number depicted in Fig.~\ref{fig:app}(a). By way of comparison, we also show in (b) the LDoS for a semi-infinite geometry along the $\bm{a}_2$ direction, evaluated at $n_2 = 1$ and $\epsilon = \hbar\Omega/2$, as a function of $k_1$, with $k_3 = 0$. For the relevant parameter region of $\Omega$, we find a perfect agreement between the number of chiral states and the calculated Chern number, i.e., bulk-boundary correspondence is verified. In addition, it is possible to observe that the chirality of the localized states is determined by the laser handedness, since the inversion $\tau \rightarrow -\tau$ naturally changes the sign of the Chern number. From the obtained result, we conclude that the contributions to the Chern number is $\tau$ times the number of bands crossing at the FZB. In fact, we distinguish four different regions for the Chern number, which coincide with those cases in which either the $p=1$ or $p=2$ bands cross this energy. For example, for $\hbar\Omega/2 \lesssim 0.25 \, \gamma_0$, only the $p=1$ bands can fulfill the crossing condition, so $c_1 = 2\tau$ and $c_2 = 0$. The opposite happens for $2.5 \, \gamma_0 \lesssim \hbar\Omega/2 \lesssim 3.5 \, \gamma_0$, where $c_1=0$ and $c_2=2\tau$. In this sense, we can say that the Chern number (and with it the number of localized states for a given $k_3$) signals the number of band crossings taking place at the FZB. 

Of course, the above analysis is valid under the assumption that the contributions from the stacking direction can be decomposed into normal modes, such that $k_3$ given by Eq.~(\ref{eq:cuts}) can be fairly taken as a fixed parameter. In this case, we notice that the interlayer hoppings depend on $k_3$, and therefore the Chern number can change from one normal mode to another. When adding up all contributions coming from the normal modes, the total Chern number
\begin{equation}
c_{\tf{FZB}} = \sum_{p=1}^2\sum_{n=1}^{N_3} c_{p,n},
\end{equation} 
varies in a similar way as in Fig.~\ref{fig:app}(a), but with the following differences: (1) such a quantity needs to be multiplied by $N_3$. (2) Around the transition region centered at $\hbar\Omega/2 = 0$, the total Chern number varies in a staggered way from $2 \tau N_3$ to $4 \tau N_3$, while around $\hbar\Omega/2 = 3 \gamma_0$ this number changes from $4 \tau N_3$ to 0. This behavior is shown in Fig.~\ref{fig:app}(c), where we calculate the number of bands that cross at the FZB as a function of the driving frequency for $N_3 = 5$ (solid red) and $N_3 = 10$ (solid blue).

We recognize that, in order to deal with a semi-analytic calculation for the Chern number, we worked in a simplified model of graphite where next-nearest-neighbor couplings and energy shifts between inequivalent carbon atoms were disregarded. The inclusion of these terms would only complicate such a calculation, though the main result would remain the same, namely, each band crossing at the FZB contributes with a factor $2\tau$ to the total Chern number. With this in mind, we only expect some differences in the number of crossings near the transition regions of Fig.~\ref{fig:app} (red shaded areas), as the bands experience slight modifications when including these terms. However, for the considered frequency value $\hbar\Omega = 2.2$ eV we used along this work, the Chern number would remain the same regardless of the value of $k_3$, so for finite samples along the stacking direction we expect $c_{\tf{FZB}} = 4\tau N_3$.

\section{Normal mode decomposition} \label{sec:app_modes}

In this section we discuss the employed normal mode decomposition in the calculation of the local density of states shown in Fig.~\ref{fig:4}. Let us consider the bulk Hamiltonian $\hat{\mathcal{H}}_{\bm{k}}$ of Eq.~(\ref{eq:HB}) whose matrix representation is given in Eq.~(\ref{eq:bloch}). We first break translational invariance along the $z$-direction, by considering $N_3$ unit cells along $\hat{\bm{a}}_3$. The set of allowed values for $k_3$ is no longer a continuum, and we expect some discrete set which we propose to be given by
\begin{equation}
k_3 = \frac{n\pi}{c_0(2N_3+1)}, \qquad n=1,2,\dots,N_3 . \label{eq:cuts}
\end{equation}
In this way, we obtain that the functions that depend on $k_3$ take the following values:
\begin{equation}
f_4 = 1 + e^{2i\varphi_n}, \text{ and  } f_5 = 2 \cos(2\varphi_n),
\end{equation}
where $\varphi_n= n\pi/(2N_3+1)$. Now, for every value $n$ and fixed $\bm{k} = (k_x,k_y)$ we can diagonalize $\hat{\mathcal{H}}_{\bm{k}} \rightarrow \hat{\mathcal{H}}_{\bm{k}}^n$, where the superscript indicates that $k_3$ is given by $n$. This yields 4 eigenenergies and their corresponding eigenkets, i.e.
\begin{equation}
\hat{\mathcal{H}}_{\bm{k}}^n \ket{\phi_{\alpha,\bm{k}}^n} = \epsilon_{\alpha,\bm{k}}^n \ket{\phi_{\alpha,\bm{k}}^n}, \qquad \alpha = 1,\dots,4,
\label{eq:eigen}
\end{equation}
where the eigenket can be written in terms of the site basis $i = \{\tf{A}_1,\tf{B}_1,\tf{A}_2,\tf{B}_2\}$ as 
\begin{equation}
\ket{\phi_{\alpha,\bm{k}}^n} = \sum_i \phi_{\alpha,\bm{k}}^n(\bm{\delta}_i) \ket{i},
\end{equation}
and $\phi_{\alpha,\bm{k}}^n(\bm{\delta}_i) = \braket{\bm{\delta}_i|\phi_{\alpha,\bm{k}}^n}$. What we do now is to translate these coefficients into a new space of dimension $4N_3$, given by the amount of units cells spanned along $\hat{\bm{a}}_3$. This is accomplished by transforming these coefficients as follows:
\begin{widetext}
\begin{equation}
\phi_{\alpha,\bm{k}}^n(\bm{\delta}_i,n_3) = \frac{2}{\sqrt{N_3+1}} \phi_{\alpha,\bm{k}}^n(\bm{\delta}_i) \left\{ 
\begin{array}{ll}
\sin\left[\dfrac{(2n_3-1)n\pi}{2N_3+1}\right]e^{-i\varphi_n/2}, & i \in \text{LL}\\
& \\
\sin\left[\dfrac{ 2n_3   n\pi}{2N_3+1}\right]e^{+i\varphi_n/2}, & i \in \text{UL}
\end{array} \right. \equiv \phi_{\alpha,\bm{k}}^n(\bm{\delta}_i) \left\{ 
\begin{array}{ll}
a_{n,n_3}, & i \in \text{LL}\\
& \\
b_{n,n_3}, & i \in \text{UL}
\end{array} \right.,
\end{equation} 
\end{widetext}
where the index $n_3 = 1,\dots,N_3$ denotes the unit cell in the finite system. We can therefore construct the following states in this new space as
\begin{equation}
\ket{\Phi_{\alpha,\bm{k}}^n} = \sum_{n_3=1}^{N_3} \sum_i \phi_{\alpha,\bm{k}}^n(\bm{\delta}_i,n_3) \ket{i,n_3}. \label{eq:ext-phi}
\end{equation}
The idea then is to test such a transformation in the full Hamiltonian that arises when translational invariance along $\hat{\bm{a}}_3$ is broken. In terms of the $\{\ket{n_3}\}$ basis, this Hamiltonian presents the following structure:
\begin{equation*}
\hat{\mathcal{H}}_{\bm{k}} = \sum_{n_3=1}^{N_3} \bm{h} \otimes \ket{n_3}\!\bra{n_3} + \sum_{n_3=1}^{N_3-1} \left( \bm{v} \otimes \ket{n_3+1}\!\bra{n_3} + \tf{h.c.} \right),
\end{equation*}
where the block matrices $\bm{h}$ and $\bm{v}$ represent the intra- and inter-cell couplings, respectively, and are defined as:
\begin{equation}
\bm{h} = \left(
\begin{array}{cccc}
\epsilon_0 + \Delta	& \gamma_0 f_1		& \gamma_1				& \gamma_4 f_2	\\
\gamma_0 f_1^*		& \epsilon_0		& \gamma_4 f_1^*		& \gamma_3 f_3	\\
\gamma_1			& \gamma_4 f_1		& \epsilon_0 + \Delta	& \gamma_0 f_2	\\
\gamma_4 f_2^*		& \gamma_3 f_3^*	& \gamma_0 f_2^*		& \epsilon_0
\end{array} \right),
\end{equation}
and
\begin{equation}
\bm{v} = \left(
\begin{array}{cccc}
\gamma_5		& 0					& \gamma_1			& \gamma_4 f_2	\\
0				& \gamma_2			& \gamma_4 f_1^*	& \gamma_3 f_3	\\
0				& 0					& \gamma_5			& 0				\\
0				& 0					& 0					& \gamma_2
\end{array} \right).
\end{equation}
If we now apply this Hamiltonian into the proposed state given by Eq.~(\ref{eq:ext-phi}),  we obtain:
\begin{equation}
\hat{\mathcal{H}}_{\bm{k}} \ket{\Phi_{\alpha,\bm{k}}^n} =  ( \epsilon_{\alpha,\bm{k}}^n \hat{\mathcal{I}} + \hat{\mathcal{V}} ) \ket{\Phi_{\alpha,\bm{k}}^n},
\end{equation}
where $\hat{\mathcal{I}}$ is the identity operator in this extended space and 
\begin{equation*}
\hat{\mathcal{V}} = \gamma_5 \left( \hat{P}_{\tf{A}_1,1} + \hat{P}_{\tf{A}_2,N_3} \right) 
		+ \gamma_2 \left( \hat{P}_{\tf{B}_1,1} + \hat{P}_{\tf{B}_2,N_3} \right),
\end{equation*}
where we defined the projectors $\hat{P}_{i,n_3} = \ket{i,n_3}\!\bra{i,n_3}$. The matrix associated with this operator is therefore diagonal, and the nonzero elements are only in the first ($n_3=1$) and last ($n_3=N_3$) unit cells. The proposed decomposition scheme, therefore, is not exact due to the next-nearest-neighbor hoppings $\gamma_5$ and $\gamma_2$ appearing in $\hat{\mathcal{V}}$. However, as both $\gamma_2$ and $\gamma_5$ are much smaller than $\gamma_0$, the operator $\hat{\mathcal{V}}$ can be taken as a small perturbation on $\hat{\mathcal{H}}_{\bm{k}}$ when we increase $N_3$, such that it can be disregarded in a first approximation. This implies that the energies $\epsilon_{\alpha,\bm{k}}^n$, obtained from a $4 \times 4$ Hamiltonian matrix, are in fact a good approximation to the exact eigenenergies, which would be obtained from a $4 N_3 \times 4 N_3$ matrix.

It is important to notice that the above presented decomposition can be extended straightforwardly to incorporate the circularly polarized light. What changes in this case is that the static Bloch Hamiltonian in Eq.~(\ref{eq:eigen}) should be replaced by the Bloch-Floquet Hamiltonian of Eq.~(\ref{eq:HamF}), and the state $\ket{\phi_{\alpha,\bm{k}}^n}$ is now defined in the $\mathcal{F}$-space, whose dimension is $4 (2n_r+1)$ (recall that $2n_r+1$ is the amount of considered Floquet replicas). Additionally, for the LDoS of Fig.~\ref{fig:4}, translational invariance is not only broken along $\hat{\bm{a}}_3$, but also in $\hat{\bm{a}}_2$. For a given $k_3$, specified by $n$, we can calculate an effective local density $\mathcal{N}_{\bm{r},n}$ by following the decimation procedure discussed in detail in Ref.~[\onlinecite{calvo2013}]. This procedure consists in the recursive calculation of the self-energy correction on the site located at $\bm{r}_0 = n_2 \bm{a}_2 + \bm{\delta}_i$, due to the presence of the other sites in the lattice. Once we obtain $\mathcal{N}_{\bm{r},n}$, the final LDoS at site $\bm{r} = \bm{r}_0 + n_3 \bm{a}_3$ can be obtained as:
\begin{equation}
\mathcal{N}_{\bm{r}} = \sum_n \mathcal{N}_{\bm{r}_0,n} \left\{ 
\begin{array}{ll}
|a_{n,n_3}|^2,	& i \in \text{LL}	\\
& \\
|b_{n,n_3}|^2,	& i \in \text{UL}
\end{array} \right. .
\end{equation}
The relevance of this decomposition scheme relies on the fact that it effectively reduces the dimension of the involved Hamiltonians, and thus the computation time demanded by the calculation of either the system's eigenenergies or the LDoS. This scheme, in turn, yields a very good approximation to the exact solutions for large values of $N_3$, such that surface effects due to the perturbation $\hat{\mathcal{V}}$ can be neglected. It is precisely in this limit where the exact calculation becomes highly demanding and, in most of cases, almost impossible to carry out.

\bibliographystyle{apsrev4-1_title}
\bibliography{cite}

\end{document}